\documentclass[]{aa}

\usepackage{txfonts}
\usepackage{graphicx}

\begin{document}

\title{Satellite content and quenching of star formation in galaxy groups at $z\sim1.8$}

\author{R. Gobat\inst{1,2}
\and E. Daddi\inst{2}
\and M. B\'{e}thermin\inst{3}
\and M. Pannella\inst{4}
\and A. Finoguenov\inst{5}
\and G. Gozaliasl\inst{5}
\and E. Le Floc'h\inst{2}
\and C. Schreiber\inst{2}
\and V. Strazzullo\inst{4}
\and M. Sargent\inst{6}
\and T. Wang\inst{2}
\and H.S. Hwang\inst{1}
\and F. Valentino\inst{2}
\and N. Cappelluti\inst{7}
\and Y. Li\inst{8}
\and G. Hasinger\inst{8}
}

\institute{
School of Physics, Korea Institute for Advanced Study, Hoegiro 85, Dongdaemun-gu, 
Seoul 130-722, Republic of Korea
\and Laboratoire AIM-Paris-Saclay, CEA/DSM-CNRS--Universit\'{e} Paris Diderot, Irfu/Service 
d'Astrophysique, CEA Saclay, Orme des Merisiers, F-91191 Gif sur Yvette, France
\and European Southern Observatory, Karl-Schwarzschild-Str. 2, 85748, Garching, Germany
\and Department of Physics, Ludwig-Maximilians-Universit\"{a}t, Scheinerstr. 1, D-81679 
M\"{u}nchen, Germany
\and Department of Physics, University of Helsinki, Gustaf-H\"{a}llstr\"{o}minkatu 2a, FI-0014 
Helsinki, Finland
\and Astronomy Centre, Department of Physics and Astronomy, University of Sussex, Brighton, 
BN1 9QH, United Kingdom
\and INAF-Osservatorio Astronomico di Bologna, Via Ranzani 1, I-40127 Bologna, Italy
\and Institute for Astronomy, University of Hawaii, 2680 Woodlawn Drive, Honolulu, HI 96822, USA
}

\date{Received 08 April 2015 / Accepted 07 July 2015}

\abstract{
We study the properties of satellites in the environment of massive star-forming galaxies 
at $z\sim1.8$ in the COSMOS field, using a sample of 
215 galaxies on the main sequence of star formation with an average mass of $\sim10^{11}$~M$_{\odot}$. 
At $z>1.5$, these galaxies typically trace halos of mass $\gtrsim10^{13}$~M$_{\odot}$. We use 
optical-near-infrared photometry to estimate stellar masses and star formation rates (SFR) of centrals 
and satellites down to $\sim6\times10^{9}$~M$_{\odot}$. We stack data around 215 central galaxies to 
statistically detect their satellite halos, finding an average of $\sim3$ galaxies in excess of the 
background density. We fit the radial profiles of satellites with simple $\beta$-models, and compare 
their integrated properties to model predictions. 
We find that the total stellar mass of satellites amounts to $\sim68$\% of the central galaxy, while 
SED modeling and far-infrared photometry consistently show their total SFR to be $25-35$\% of the 
central's rate. We also see significant variation in the specific SFR of satellites within the halo 
with, in particular, a sharp decrease at $<100$~kpc. After considering different potential explanations, 
we conclude that this is likely an environmental signature of the hot inner halo. This effect can 
be explained in the first order by a simple free-fall scenario, suggesting that these low-mass 
environments can shut down star formation in satellites on relatively short timescales of 
$\sim0.3$~Gyr. 
}

\titlerunning{Star formation in $z\sim1.8$ groups}
\authorrunning{Gobat et al.}

\maketitle

\section{Introduction}

Although the gradual infall of small dark matter halos onto larger ones has become a relatively 
straightforward aspect of the standard hierarchical formation paradigm, what happens to the baryons 
they contain is less well understood. In particular, the mechanisms that drive the evolution of their 
constituent galaxies become more complex as they are accreted by larger structures. Of special 
relevance are the processes that regulate and ultimately suppress star formation in galaxies in the 
early Universe.  Their relationship to, and influence on, the galaxies' immediate environment is 
not known with certainty; also debated is the relative importance of internal mechanisms versus 
externally driven ones
\citep[although the former are expected to be dominant in massive systems and the latter to act 
preferentially on lower-mass satellite galaxies; e.g.,][]{Bal06,Pen10,Gab11}.
The $z=1.5-2.5$ epoch is particularly interesting as a transition period when global 
star formation in the universe peaks, but also where the first ostensibly 
collapsed and virialized galaxy structures appear, in which a spatial segregation of 
different galaxy types (e.g., passive and active) is observed. 
In particular, the cores of massive clusters appear to become dominated by quiescent galaxies 
around this time \citep[e.g.,][]{Spi12,Stra13,Gob13}. From a theoretical point of view, 
the increasing temperature of the gaseous medium in group- and cluster-scale halos starts 
to efficiently prevent accretion around this epoch \citep[e.g.,][]{Dek06,Dek09}, thus affecting 
the build-up and evolution of the galaxies they host. One can therefore expect the processes 
regulating mass accretion onto galaxies, crucial to our understanding of galaxy build-up, to be 
relatively accessible to observation at this epoch.\\

For practical and historical reasons, the mass regime most often explored at high 
redshift has been that of large galaxy clusters, which are the richest and most readily 
selectable halos. They are also the most biased regions in which to study environmental effects 
on galaxy evolution. However, at high redshift the advantages offered to galaxy evolution studies 
by their galaxy density and mass contrast are somewhat counterbalanced by their relative rarity. 
A lot of effort has thus been devoted to the search for high-redshift structures, 
through a variety of methods. Although remarkable progress has been made recently, only 
a handful of $z>1.5$ structures have been accurately characterized so far 
\citep[e.g.,][]{And09,Pap10,Gob11,Stan12}, as this endeavor is still fundamentally hampered 
by the relatively limited area for which deep datasets are available (although this may 
change thanks to ongoing and future deep wide-field infrared surveys). 
On the other hand, the lower ``group'' mass range, at this redshift that of the progenitors 
of $z=0$ clusters, has been less systematically explored; it requires either much deeper 
data \citep[e.g.,][]{Erf13,Tan13} or tracers, such as specific galaxy types (e.g., quasars 
or giant radio-galaxies), that correlate with structures but are not directly proportional 
to mass (unlike, e.g., galaxy density or diffuse X-ray emission). This latter type of tracer has 
been historically used for higher redshift systems, such as proto-clusters 
\citep[e.g.,][]{Ha11,Wyl13}. At lower halo masses, even isolated massive galaxies are expected 
to be at the center of galaxy assemblages, as a simple consequence of the hierarchical nature 
of matter distribution in the universe.\\ 

In \citet{Bet14}, we indeed found that massive ($\sim10^{11}$~M$_{\odot}$), star-forming 
galaxies on the main sequence of star formation \citep[e.g.,][]{Bri04,Dad07,Rod10} in the range 
$z=1.5-2.5$ had clustering properties consistent with halos of mass ($>10^{13}$~M$_{\odot}$). 
Subsequent stacking of deep X-ray datasets available in COSMOS yielded constraints on the total mass 
consistent with the clustering analysis ($\sim2\times10^{13}$~M$_{\odot}$), thus providing independent 
confirmation. This suggests that massive main-sequence galaxies constitute a conspicuous 
tracer of group-scale environments at $z\sim2$ (more so than, e.g., isolated quiescent galaxies), 
thus allowing for the easy study of these systems.\\

Small halos comprising a central galaxy and its satellite system are particularly useful for 
probing the environmental dependency of galaxy properties over large mass and redshift ranges. 
In particular, they provide a powerful tool to constrain quenching mechanisms and timescales, 
through their tell-tale signature on mass profiles and functions 
\citep[e.g.,][]{Wan10,Wan14,Phi14,Har15} or simple derived quantities such as the fraction 
of quiescent satellites \citep[e.g.,][]{Geo11}. 
Being vastly more abundant and structurally simpler than massive galaxy clusters, these 
systems allow for a straightforward test for galaxy assembly and evolution models 
\citep[e.g.,][]{Guo11}, without requiring deep knowledge of their components (age, position 
in phase-space, etc.). 
Here we have taken an intermediate approach, focusing on the properties of star-forming 
satellites, and in particular on the variation of their star formation rates (SFR).
We have only considered systems with massive star-forming centrals: while quiescent galaxy 
pairs also trace similar sized halos \citep{Bet14}, their center of mass is less clear, 
which would make an investigation of the radial dependency of satellite properties 
less straightforward.
This paper is organized as follows: in Section \ref{data}, we describe the dataset and our 
sample selection. In Section \ref{stack}, we present the integrated properties of satellites 
and the method used to derive them. In Section \ref{sfr}, we discuss their variation with 
environment and present our conclusions in Section \ref{conc}. Throughout this paper we assume 
a $\Lambda$CDM cosmology with $H_0=70$~km~s$^{-1}$~Mpc$^{-1}$, $\Omega_m=0.27$, and 
$\Lambda=0.73$, and a \citet{Cha03} initial mass function (IMF; relations used here that 
assume a different IMF have been converted to this one). Magnitudes are given in the AB 
photometric system throughout.

\section{\label{data}Data and sample selection}

In this work, we have used aperture-corrected photometry from the $K_s$-selected catalog of the 
COSMOS/UltraVISTA survey \citep{McC12} from \citet{Muz13}. We have adopted photometric redshifts 
($z_{phot}$) from \citet[and references therein]{Il13} rather than the $z_{phot}$ estimates from 
\citet{Muz13}, as the former had access to a larger and deeper training set of spectroscopic redshifts 
($z_{spec}$) \citep[especially zCOSMOS Deep;][]{Lil07}, ensuring greater reliability of $z_{phot}$ at 
$z>1$. 
For objects for which these were not available, we have used the $z_{phot}$ estimates from 
\citet{Muz13}. Where possible, we have also used $z_{spec}$ from the zCOSMOS Bright sample 
\citep{Lil09}. Stellar masses, SFR, and rest-frame colors were then recomputed based on this 
merged catalog, as described in Section \ref{spm}. Finally, we have also used mid- and 
far-infrared (FIR) maps from \emph{Spitzer}/MIPS \citep{LeF09}, \emph{Herschel}/PACS 
\citep[from the PEP survey;][]{Lut11}, and \emph{Herschel}/SPIRE 
\citep[from the HerMES survey;][]{Oli12} in the analysis, although only the PACS data were 
used for the construction of the sample (see below).\\

\begin{figure}
\centering
\includegraphics[width=0.49\textwidth]{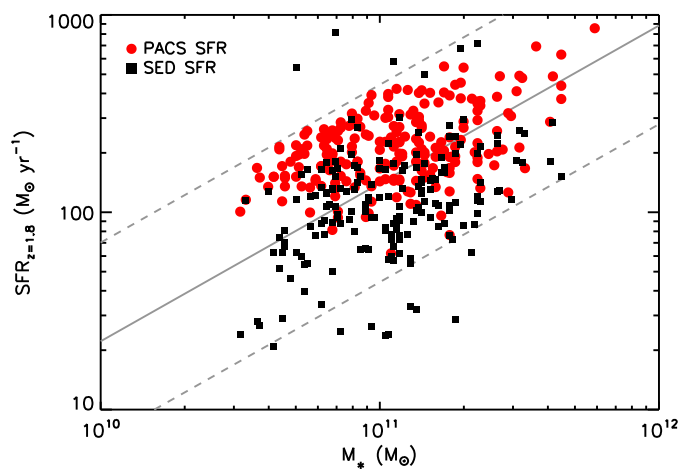}
\caption{SFR of centrals as a function of their stellar mass, for SFR estimates from, respectively, 
\emph{Herschel}/PACS fluxes (red circles) and SED modeling (black squares). Our adopted parametrization 
of the main sequence at $z=1.8$ is shown as a gray line, with the 0.5~dex limit shown by dashed lines. 
The SFRs shown here have been corrected according to the difference in normalization between the main 
sequence at $z=1.8$ and the redshift of the objects.}
\label{fig:ms}
\end{figure}

\begin{figure*}
\centering
\includegraphics[width=0.33\textwidth]{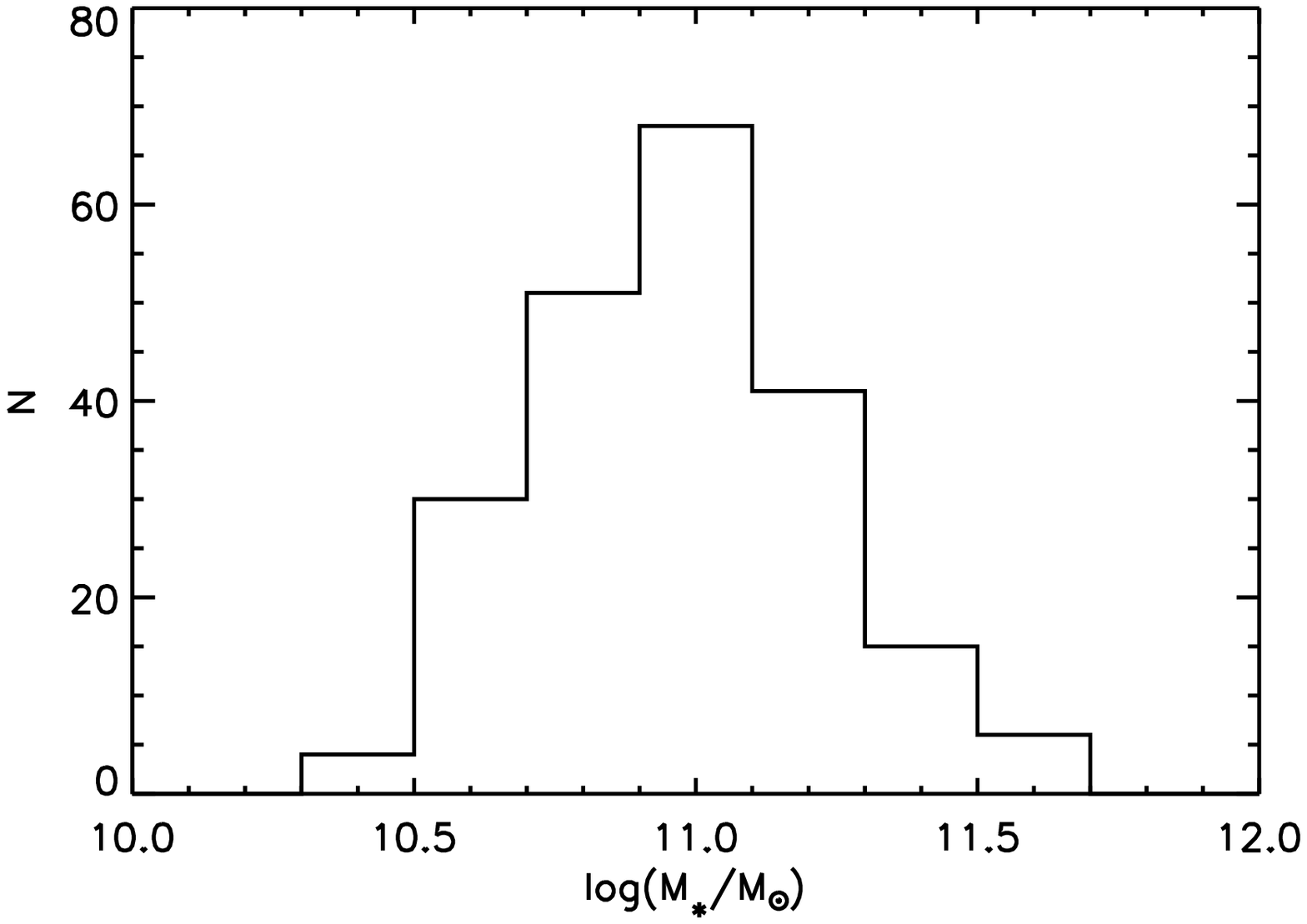}
\includegraphics[width=0.33\textwidth]{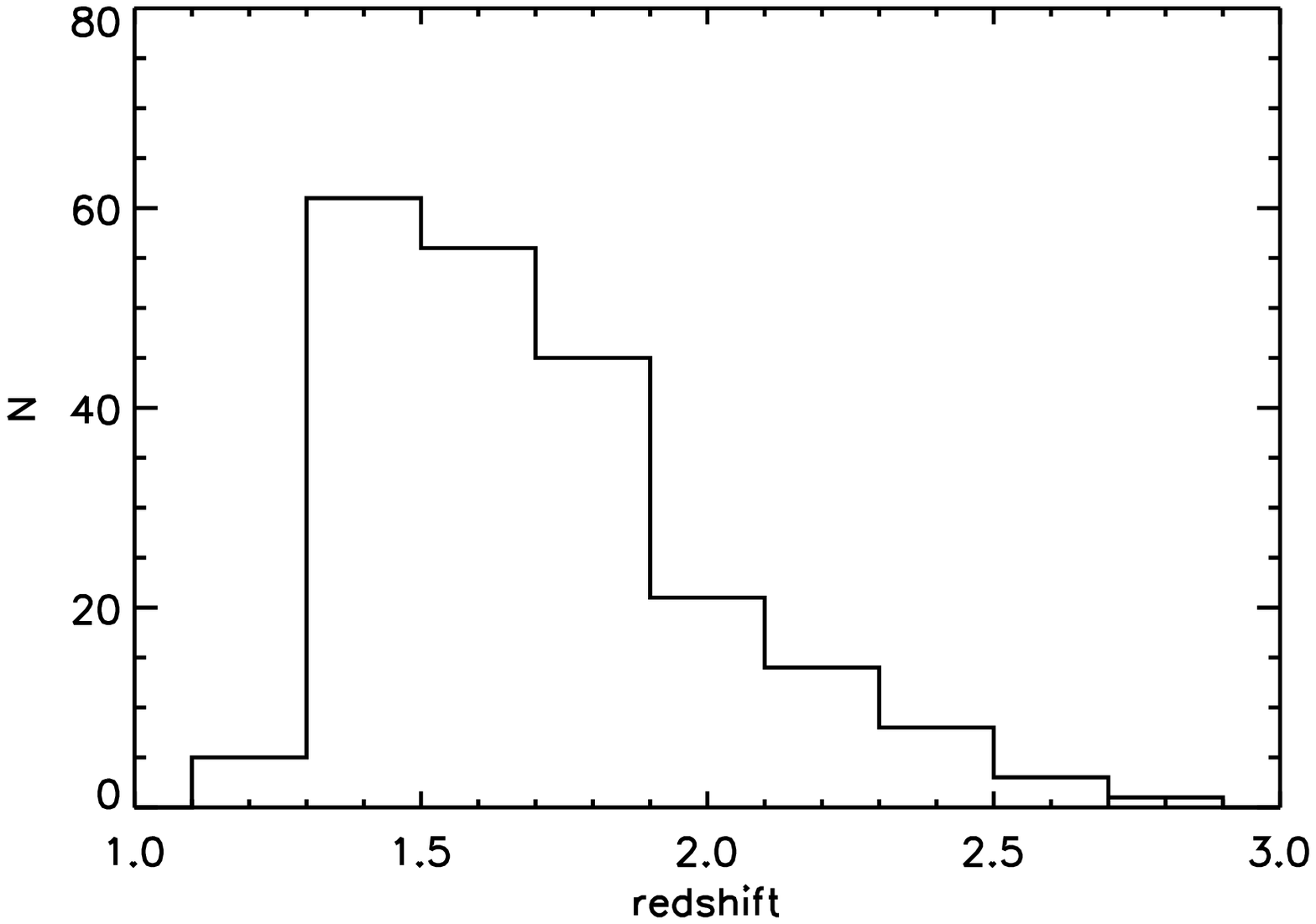}
\includegraphics[width=0.33\textwidth]{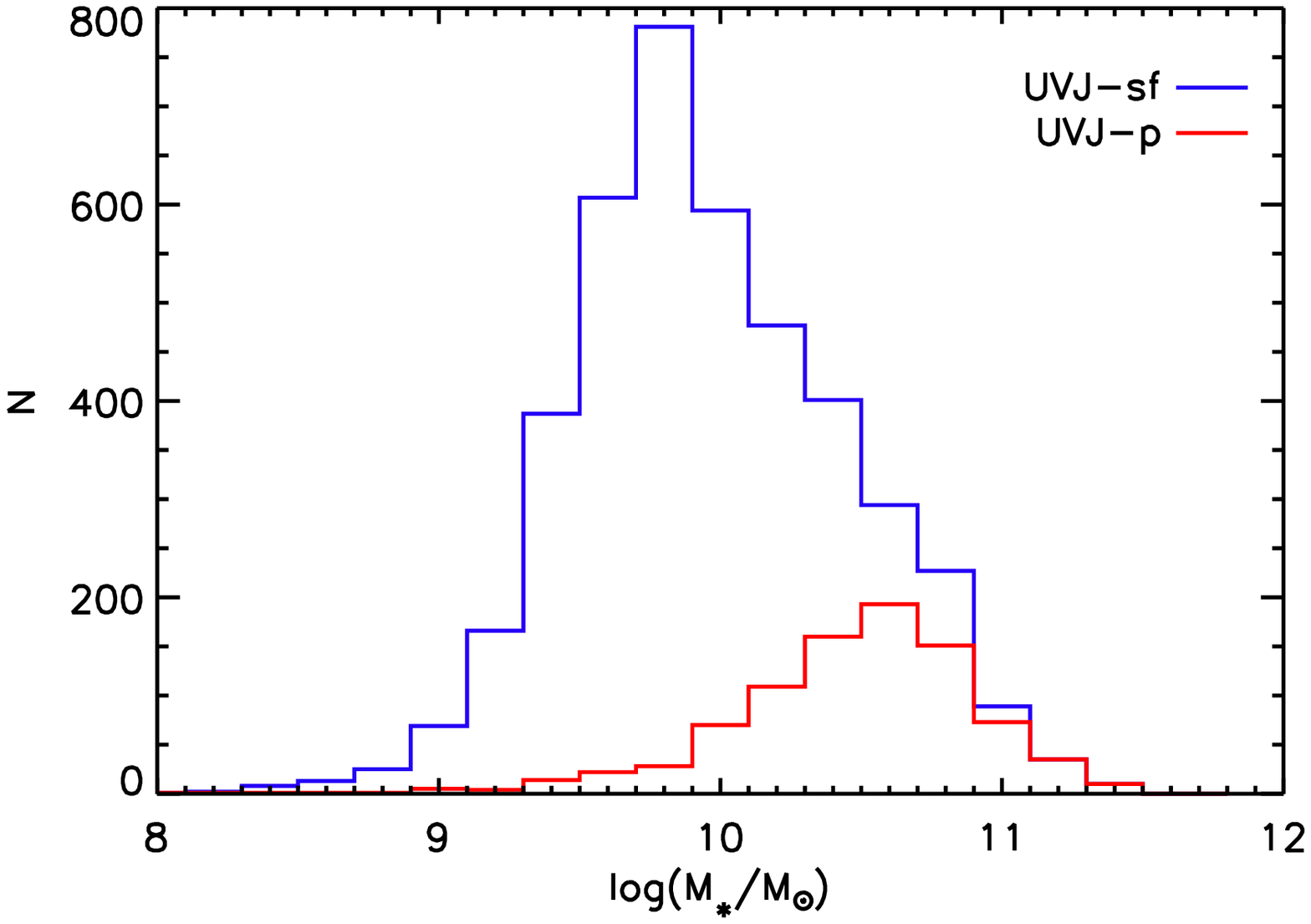}
\caption{Distribution of stellar masses (left) and redshifts (middle) of centrals 
in our sample, and of stellar masses of satellites (right) color-selected as 
star-forming (blue histogram) and quiescent (red histogram).}
\label{fig:mz}
\end{figure*}

For consistency with \citet{Bet14}, we have built a sample of massive, star-forming galaxies by 
considering all BzK-selected \citep{Dad04}, \emph{Herschel}/PACS-detected objects with SFRs 
within 0.5~dex of the main sequence \citep[as parametrized in][]{Bet12}, using the recomputed stellar 
mass derived from the UV-near-infrared (NIR) photometry and SFR derived from fits to the $100$ and 
$160\mu$ fluxes with \citet{Mag12} templates. 
Selecting only PACS-detected sources mostly yields massive galaxies, while the 0.5~dex removes 
lower and upper outliers (e.g., quenching galaxies and starbursts, respectively). The position of 
galaxies in our sample, relative to the main sequence at $z=1.8$, is shown in Fig.~\ref{fig:ms}. 
In addition, we have also rejected galaxies that are less than 1\arcmin~($\sim500$~kpc) 
away from known overdensities with redshifts consistent within the 68\% confidence level 
\citep{Chi14,Stra15}, or probable companions (also with redshifts consistent at 68\% 
confidence) of mass $m_{frac}\geq2$ times higher within $r_{vir}=35$\arcsec~($\sim300$~kpc).
This constraint corresponds roughly to the typical expected virial radius ($r_{vir}$) of a 
$\sim2\times10^{13}$~M$_{\odot}$ halo in the sample's redshift range ($1.5\lesssim z\lesssim2.5$), 
while the constraint on the mass ratio between the central and companion galaxy reflects the 
typical stellar mass uncertainty when all variables and degeneracies are taken into account 
\citep[e.g.,][]{Ber04,Con13}. The first criterion is conservative and meant to minimize the 
risk of including halos significantly more massive than $\sim2\times10^{13}$~M$_{\odot}$. 
Such systems would also be richer and might somewhat bias the results of our analysis.
These values, although physically motivated, were also chosen as a compromise to obtain a relatively 
clean but still statistically significant sample. We note that altering them slightly (e.g., 
$m_{frac}=1$~and $r=5$\arcsec~so as to reject probable interacting pairs) does not change 
the sample much, nor the outcome of the analysis presented below. 
This criterion yields 215 galaxies, which we henceforth assume are central to their halo (hereafter, 
centrals), with a mean mass of $\langle$M$_{\star}\rangle=1.3\times10^{11}$~M$_{\odot}$ and a mean 
redshift of $\langle z\rangle =1.8$. The scatter of the centrals' stellar mass distribution is 
$\sim0.25$~dex, consistent with that expected from a single halo population at this redshift 
\citep{Beh13}. The distribution of masses and redshifts in the sample is shown in Fig.~\ref{fig:mz}.\\

\subsection{\label{spm}Stellar population modeling}

We have estimated stellar masses, SFR, and dust extinction for both centrals and satellites 
(see Section~\ref{stack}) by fitting the available UV-NIR SEDs with two different types 
of \citet{BC03} models: 
stellar masses were computed assuming a generalistic delayed exponential star formation 
history (SFH) of the form SFR(t)~=~SFR$_0\times (t/\tau^2)\times \textrm{exp}(-t/\tau)$ which 
includes both rising and declining cases. The age $t$ and timescale $\tau$ were allowed to 
vary between 0.1~Gyr ($t$) or 1~Myr ($\tau$) and the age of the Universe at the redshift 
of the galaxy. Star formation and extinction values, on the other hand, were estimated by 
fitting the rest-frame UV photometry assuming a constant SFR with an age limit of $t>100$~Myr
\citep[the timescale explicitly assumed for UV-derived SFRs; e.g.,][]{Ken98}. 
This was done because, while the rest-frame optical-NIR part of the SED reflects the entire 
SFH of the galaxy, the rest-frame UV part is sensitive to ``instant'' SFR. This also makes 
comparison with the literature easier, since direct UV-to-SFR conversions \citep{Ken98} 
generally assume continuous star formation over $\sim100$~Myr (see also Section \ref{sfr}). 
We have included extinction by dust, considering values of $E(B$-$V)\leq2$ and assuming a 
\citet{Cal00} functional form with an additional UV bump \citep{Nol09,Bu12}. 
This limit should be safely above the normal values for even the most massive centrals in 
our sample \citep[e.g.,][]{Gar10,Zah14,Pan14}. Solar metallicity was assumed for all models: 
due to the age-metallicity degeneracy, this parameter would only matter in the 
case of very old populations, which are not liable to be relevant for the galaxies we consider 
in our analysis.
We have ignored the far- and near-UV GALEX bands, as the resolution of GALEX is significantly 
poorer than that of the other instruments contributing to the catalog ($>4$\arcsec~compared 
to sub-arcsecond seeing), which precludes efficient deblending on scales typical of the size 
of halo cores ($\sim5$\arcsec, and thus relevant to the analysis in Section \ref{sfr}). 
For the same reason, we did not include the 5.8 and 8~$\mu$m \emph{Spitzer}/IRAC bands in 
the modeling and treated the FIR data separately, as detailed in Section \ref{ir}. The 
SED modeling was thus performed on a maximum of 25 photometric bands.\\

\section{\label{stack}Stacking analysis and integrated properties}

For each central galaxy, we have constructed a sample of candidate members of that galaxy's halo 
(hereafter ``satellites'') by selecting all uncontaminated, non-stellar sources in the merged 
catalog with $K_s<23.4$, the completeness limit cited by \citet{Muz13}, and with 
$z_{L68}<z_{cen}<z_{H68}$. Here $z_{cen}$ is the photometric redshift of the central and 
$z_{L68}$ ($z_{H68}$) the lower (respectively upper) 68\% confidence limit to the $z_{phot}$ of 
the putative satellites. The uncertainty on the association between galaxies is likely 
to be dominated by the uncertainty on the photometric redshifts of the fainter satellites, rather 
than of the bright centrals. Accordingly, we assume the redshift of the centrals to be fixed at 
the best-fit value.
In addition, we have only considered sources within a radius of 80\arcsec, or $\sim700$~kpc, about 
twice the typical $r_{vir}$ of the host halos at this redshift. Since the average minimum separation 
between centrals in our sample is $\sim2$\arcmin, and the redshift range in which satellites are 
selected can be large, this has allowed us to estimate the contribution of background and foreground 
interlopers while minimizing cross-contamination between satellite systems. 
These satellite subsamples were then decomposed into passive and star-forming galaxies based on 
their rest-frame $U$-$V$ and $V$-$J$ colors \citep[e.g.,][hereafter UVJ]{Wuy07}.
The mass distribution of satellites selected as star-forming and passive, respectively, is shown 
in  Fig.~\ref{fig:mz}. The distribution of passive-selected satellites is shown for completeness 
only, as we focus on star-forming systems from Section \ref{ir} on.\\ 

\begin{figure*}
\centering
\includegraphics[width=0.45\textwidth]{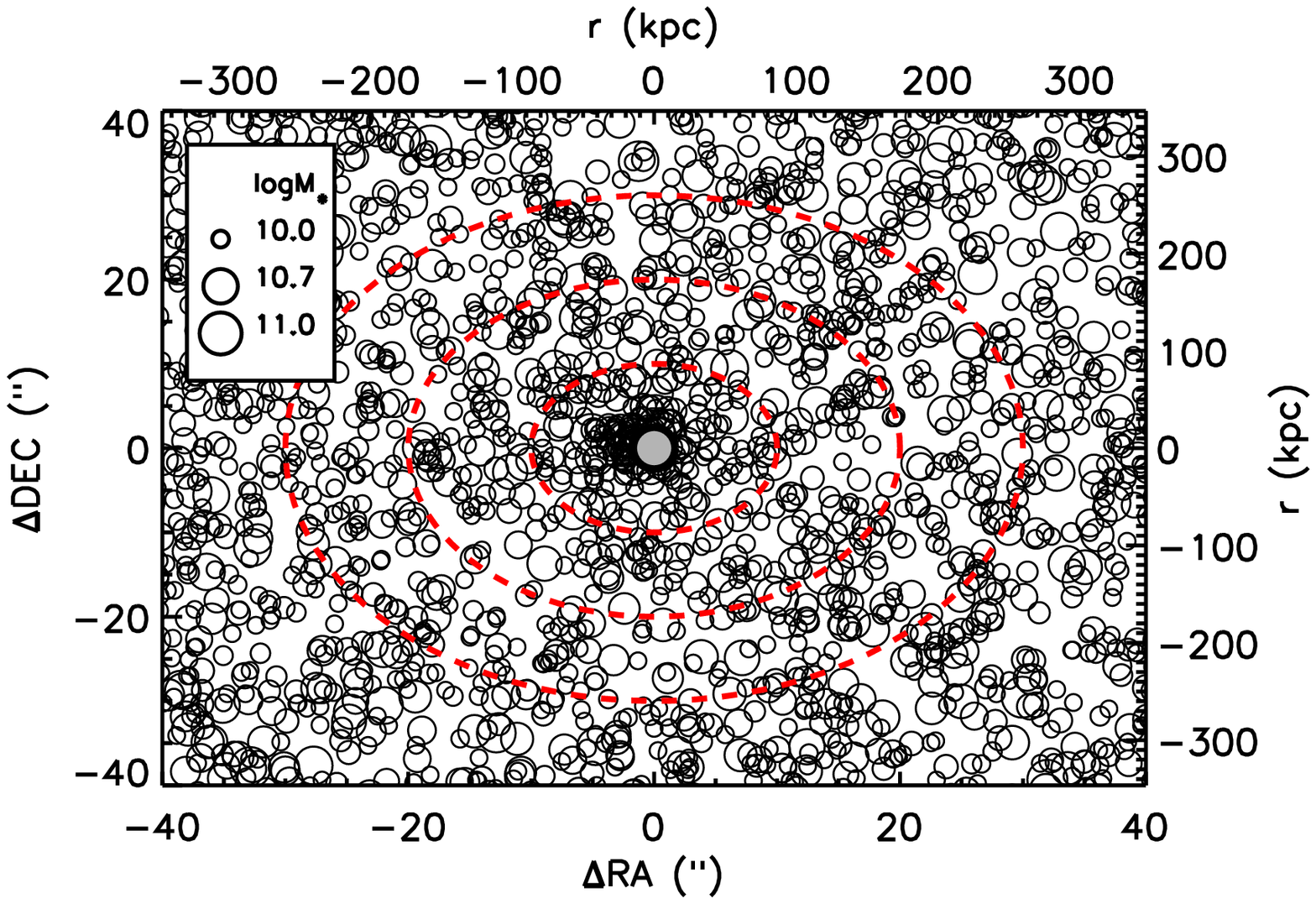}
\includegraphics[width=0.45\textwidth]{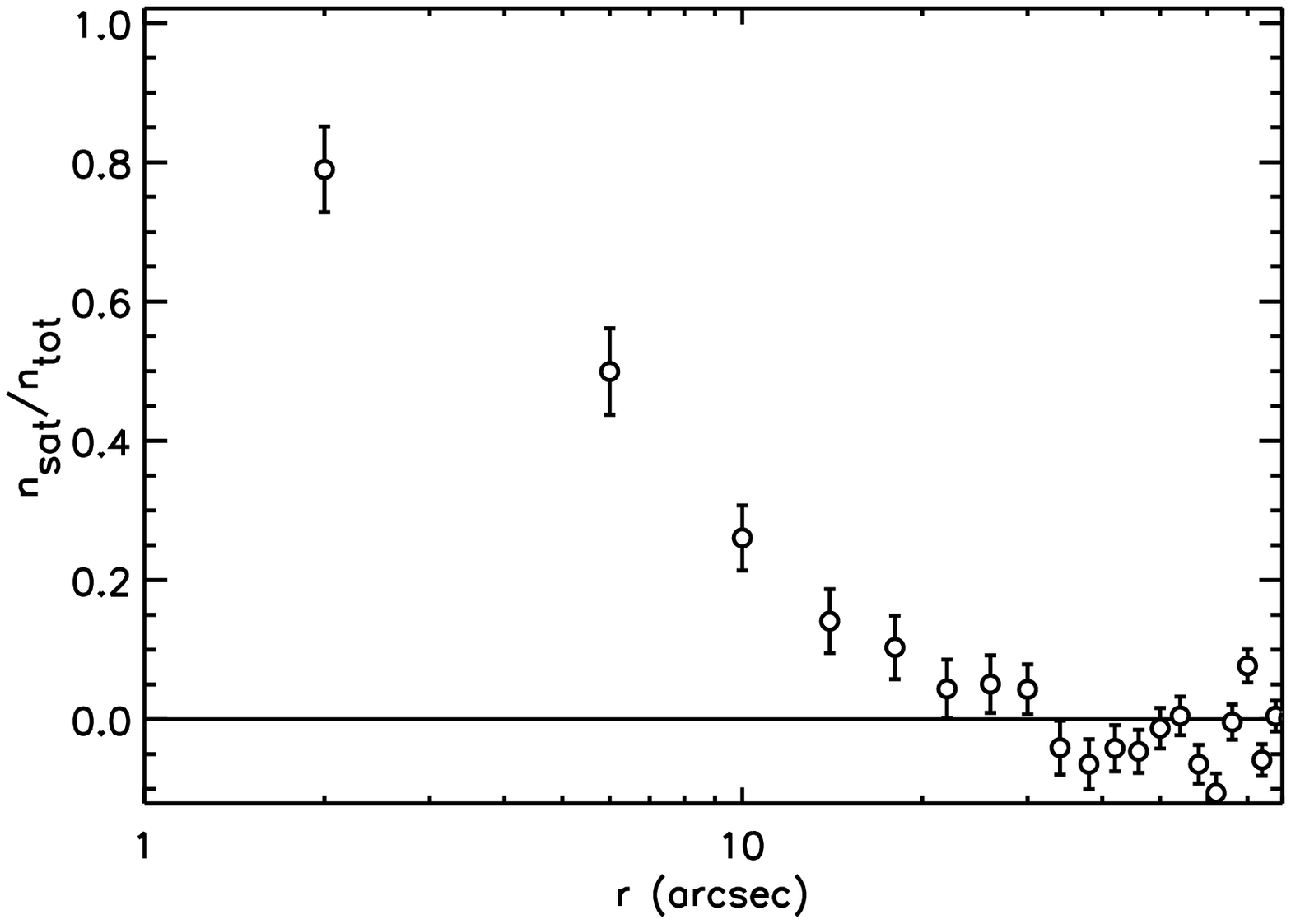}
\caption{\emph{Left:}~Positions of all satellite candidates (open circles) compared to 
their respective centrals (filled circle), with the overdensity most visible at 
$\lesssim10$\arcsec. The size of each symbol varies as a function of its stellar mass. 
The red dashed contours show scales of 10\arcsec, 20\arcsec, and 30\arcsec, respectively. 
\emph{Right:}~fraction of expected ``real associations'' among satellite candidates, 
$n_{sat}/n_{tot}$ as a function of radius (where $n_{sat}(r)=n(r)-n_b$, $n(r)$ and $n_b$ 
being, respectively, the number density of satellite candidates at radius $r$ and at 
$>50$\arcsec).}
\label{fig:pos}
\end{figure*}

\begin{figure*}
\centering
\includegraphics[width=0.8\textwidth]{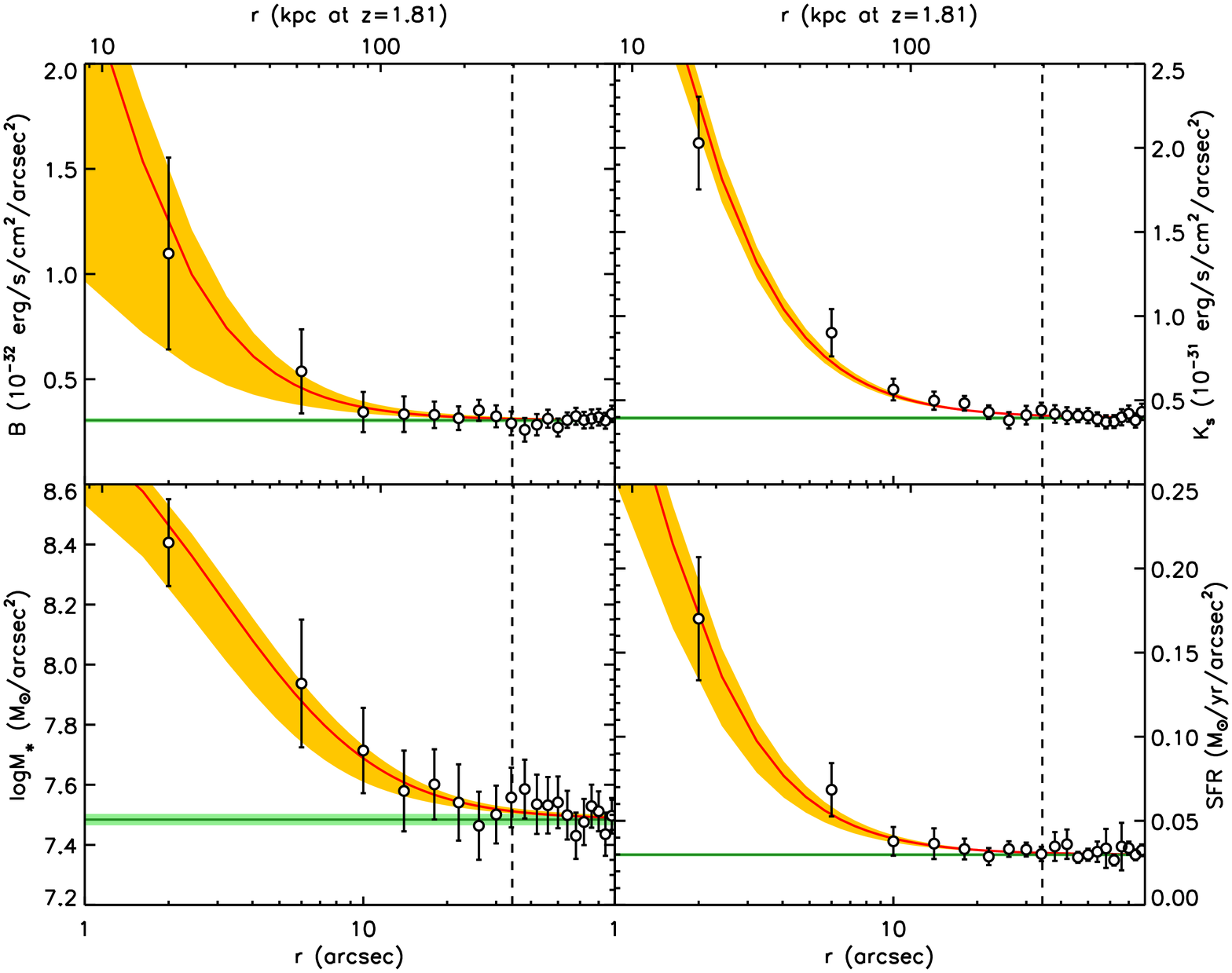}
\caption{Density profiles of satellites, shown in bins of 4\arcsec~($\sim35$~kpc 
at $z\sim1.8$) for clarity. Clockwise from the top left: $B$- and $K_s$-band flux, SFR and 
stellar mass.  The best-fit $\beta$-model, shown with its associated uncertainties 
by the red line and orange envelope, was derived from the unrebinned data. The dashed vertical 
line marks the average putative virial radius of the halos and the level of contamination by 
interlopers (i.e., ``background'') is shown in green. 
}
\label{fig:prof}
\end{figure*}

Satellites were grouped in concentric annuli of width 2\arcsec~($\sim17$~kpc) centered on 
each halo galaxy, and their relevant properties (photometric fluxes, stellar masses and 
SFR) averaged in each radial 
bin. These values were then divided by the total area of the annulus.
In each case, the contribution of background and foreground interlopers (hereafter, 
``background'', for convenience) was estimated from the values in annuli between 
50\arcsec~and 80\arcsec. This region was chosen to be comfortably distant from any 
significant galaxy excess due to the host halo (as seen in Fig.~\ref{fig:pos}) and 
cover a large area. We note that this method yields values consistent with background 
levels determined through random apertures. However, the use of an annular region that is 
still relatively close to the central should better account for local background variations 
due to interloper clustering (e.g., galaxy filaments). 
We use the same background radii for all subsamples and do not attempt to, e.g., adapt the 
size of the bins to the mass of the central, since $r_{vir}$ is not very sensitive to halo 
mass variations and the stellar mass scatter of the centrals is consistent with that of a 
single halo population.
Uncertainties in stacked properties were estimated, in each bin, using 1000 bootstrap 
resamplings of the data, with sizes of half the initial sample. 
However, bootstrap-derived uncertainties tend to be underestimated as they do not 
account for systematic uncertainties. This is typically more noticeable in the case of 
large samples. In an attempt to compensate for it, we rescaled the bootstrap estimates so 
that, when fitting the background as a constant term, the reduced chi-square be $\chi^2_0=1$ 
if initially larger. This is an ad hoc correction meant to produce more conservative 
error estimates (however, systematic errors are not necessarily Gaussian and the use of a 
$\chi^2$ estimate might not be formally justified). 
As an example, Fig.~\ref{fig:prof} shows the averaged stellar mass, SFR, $B-$, and $K_s-$band 
flux density profiles of satellites.\\

The average number of satellites per central, and the contribution of satellites to the total 
stellar mass and SFR of the halos, were estimated by fitting a parametric function to the 
profiles. We have considered both the NFW profile \citep{nfw} and projected $\beta$-model 
\citep{bmod}, commonly used for, respectively, dark matter halos and galaxy clusters. We find 
that the latter provides significantly better agreement ($\Delta$AIC$\sim45$) with 
the data, especially at small ($r<5$\arcsec) radii where the NFW profile is too shallow. 
This might be the result of mild mass segregation, as shown in Section~\ref{sfr} 
\citep[see also, e.g.,][]{Wat12,vdB14,Pis15}.
On the other hand, the profiles are well fit with $\beta\sim0.9$. Although a precise constraint 
of the true shape of the satellite profile is beyond the scope of this paper, this is consistent 
with values derived at both low and high redshift \citep[e.g.,][]{Pop04,Stra13}. 
As a check, we have also performed fits letting the background vary freely, which yielded 
background values consistent with those estimated from the outer annuli.\\ 

Integrating the number density profile, we find the average excess of $K_s<23.4$ satellites 
to be $3.3\pm0.2$. This is somewhat above the value reported by \citet{Har15} for 
high-redshift centrals and could reflect the different nature of our sample as well as the 
higher average mass of its centrals. 
Similarly, we find the integrated stellar mass and SFR of satellites to be 
M$_{\star}=(3.6\pm1)\times10^{10}$~M$_{\odot}$ and SFR~$=28\pm6$~M$_{\odot}$~yr$^{-1}$, or 
respectively 28\% and 15\% of the average central mass and SFR ($1.3\times10^{11}$~M$_{\odot}$ 
and 192~M$_{\odot}$~yr$^{-1}$, respectively).
However, these values are underestimates for two reasons: first, when selecting satellite 
candidates, we have considered objects with photometric redshifts consistent with the central's 
at only the 68\% confidence level. This was done to minimize background contamination when 
estimating radial profiles and trends (see above and Section~\ref{sfr}). However, assuming 
to the first order that the width of the intrinsic velocity distribution of satellites is 
negligible compared to photometric errors, and that the background redshift distribution is 
mostly flat in the redshift range of the selection, we can expect to lose 32\% of satellites to 
redshift uncertainties. The actual integrated mass and SFR should then be higher by a factor of 1.47. 
Secondly, we have only considered satellites within the completeness limit of the catalog, 
$K_s=23.4$, and thus do not include fainter satellites. 
At $z\sim2$, the $K_s$ band does not trace stellar mass equally for all galaxies, due to the 
flux contribution from young stars becoming non-negligible. Through comparison with stellar 
population models, we have estimated the corresponding mass limits for quiescent and star-forming 
galaxies to be, respectively, logM$_{\star}$=10.3 and 9.8. The models used here were based on 
\citet{BC03} templates assuming, respectively, a single burst of maximal age and a main-sequence 
SFH of the form 
\begin{equation}
\textrm{SFR(t)} = 10^{-10.2} \times \textrm{M}_\star(t) \left 
( \frac{\textrm{M}_\star(t)}{10^{11}\, \textrm{M}_\odot}\right)^{-0.2} 
(1+z(t))^3~~\textrm{M}_{\odot}~\textrm{yr}^{-1}
\end{equation}
where $M(t)$ and $z(t)$ are, respectively, the stellar mass and redshift at time $t$ after 
the onset of star formation. 
The integrated stellar mass of UVJ-selected passive and star-forming satellites were then 
individually corrected using \citet{Tom14} mass functions (MF) extrapolated to $10^{6}$~M$_{\odot}$. 
A correction to the SFR was similarly estimated, based on the main sequence parametrization 
shown above. 
The corrected total stellar mass and SFR are then 
M$_{\star,tot,sat}=(9.2\pm2)\times10^{10}$~M$_{\odot}$ and 
SFR$_{tot,sat}=(79\pm15)$~M$_{\odot}$~yr$^{-1}$, or respectively $\sim68$\% and $\sim35$\% of 
the contribution of the central galaxy. 
Here we have used a ``canonical'' value of 0.8 for the slope of the main sequence 
\citep[e.g.,][]{Rod14}. Some recent works, on the other hand, tend to favor a value of near-unity 
\citep{Ab14,Sch15}. If we assume this value, the corrected SFR value becomes 
$\sim54\pm10$~M$_{\odot}$~yr$^{-1}$, or $\sim24$\% of the central's. We note that using 
\citet{Il13} MFs, derived on the same field but from a shallower sample, yields very similar 
values.\\

\subsection{\label{ir}Far-infrared stacks}

\begin{figure}
\centering
\includegraphics[width=0.49\textwidth]{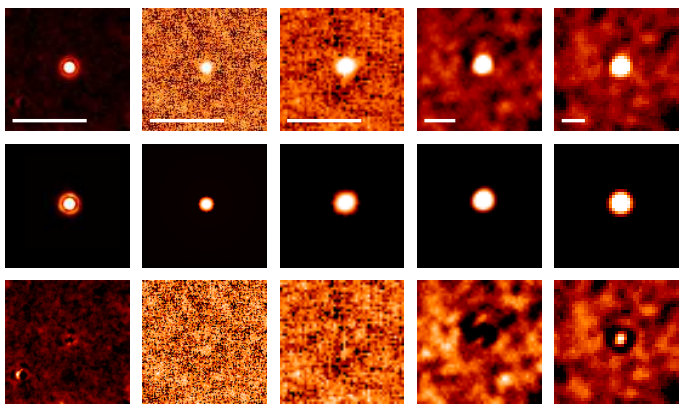}
\caption{Stacked cutouts (top), best-fit 2D models (middle), and residual images (bottom) from 
the 24$\mu$m \emph{Spitzer}/MIPS, 100 and 160$\mu$m \emph{Herschel}/PACS, 250 and 350$\mu$m 
\emph{Herschel}/SPIRE data (left to right; the 500$\mu$m SPIRE image is not shown).
The white bar in each top cutout has a length of 30\arcsec. We note that in both the cutouts 
and model images the flux is dominated by the central point-source.}
\label{fig:ir2d}
\end{figure}

On the other hand, we have extrapolated the MFs to a mass range where they are not constrained, 
and for an environment with higher density than the sample from which they were defined. 
Similarly, the slope of the main sequence at this redshift is mostly unknown below 
$\sim10^9$~M$_{\odot}$. Furthermore, extinction-corrected SED models might still underestimate 
SFRs in the case of high dust obscuration. The estimated total mass and SFR of satellites are 
therefore uncertain. 
As an independent check, we thus also estimated the total SFR of satellites from available 
FIR maps. Since the resolution of these data (FWHM$\sim6-20$\arcsec) is similar to (or even 
larger than) the characteristic size of the satellite profile, as seen in Fig.~\ref{fig:prof}, 
an annulus-based analysis would likely assign a significant fraction of the IR flux emitted 
by satellites to the central. In addition, satellites close to the mass limit are unlikely 
to be detected in the relatively shallow \emph{Herschel} maps, regardless of their separation 
from the central.
To derive the total contribution of satellites to the infrared flux of the halos, we have 
instead stacked, for each band, 80\arcsec$\times$80\arcsec~cutout images around each central. 
These stacked 2D images were then decomposed into a central point source (for the central), 
a PSF-convolved $\beta$-model centered at the same position (for the satellites) and a 
constant background term. For this fit, we have used MIPS 24~$\mu$m PSF images based on 
observations of the GOODS-North field \citep[as used in][]{Elb11} and \emph{Herschel} 
PSFs provided by the PEP and HerMES collaborations. The parameters of the $\beta$-model 
were fixed to those derived from the catalog-based SFR stack and flux uncertainties were 
estimated through Monte Carlo simulations based on parameter errors yielded by the fit, after 
renormalization so that the $\chi^2$ be at least one per degree of freedom. 
Fig.~\ref{fig:ir2d} shows the stacked images in the MIPS 24~$\mu$m, \emph{Herschel}/PACS 
100 and 150~$\mu$m, and \emph{Herschel}/SPIRE 250 and 350~$\mu$m bands, along with the best-fit 
2D models and residuals. The decomposition fails in the case of the SPIRE data, yielding only 
upper limits for the first two bands and providing no meaningful constraint to the 500~$\mu$m 
flux of satellites. This is likely due to the high confusion limit of the instrument, precluding 
a straightforward determination of the background (as shown by the residual images in 
Fig.~\ref{fig:ir2d}), and to the size of the beam being comparable to or larger than that of the 
halos themselves.\\

\begin{figure}
\centering
\includegraphics[width=0.49\textwidth]{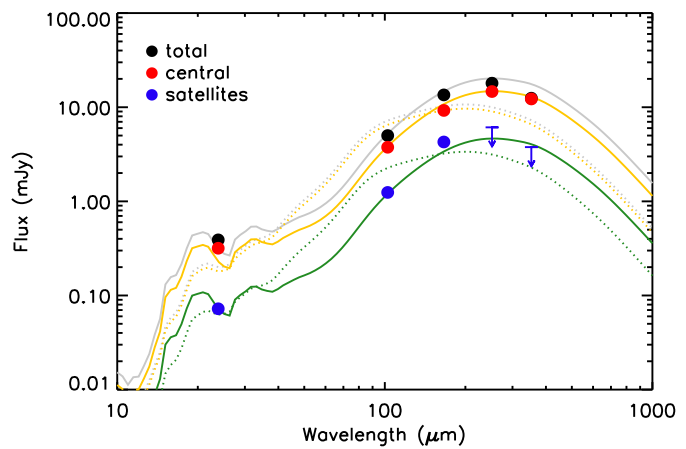}
\caption{Average far-infrared SED (\emph{Spitzer}/MIPS 24~$\mu$m, \emph{Herschel}/PACS 
100 and 160~$\mu$m, and \emph{Herschel}/SPIRE 250 and 350~$\mu$m) of centrals and 
satellites, with best-fit \citet{Mag12} main-sequence templates (solid lines). As 
comparison, best-fit starburst templates are shown as dotted lines. 
Both sets of templates have been broadened according to the centrals' redshift 
distribution.
}
\label{fig:irsed}
\end{figure}

The total infrared luminosity $L_{IR}$ was then derived from the resulting SEDs using 
\citet{Mag12} templates convolved with the redshift distribution of centrals. We have considered 
both main-sequence and starburst templates, and found that the latter perform significantly 
worse, as shown in Fig.~\ref{fig:irsed}. Converting $L_{IR}$ into SFR assuming the \citet{Ken98} 
relation, we find SFR$_{tot,IR}=176\pm11$ and $47\pm6$~M$_{\odot}$~yr$^{-1}$, for the centrals 
and satellites respectively, consistent with the values derived from extinction- and 
completeness-corrected UV SFRs (which we then use in Section \ref{sfr}). We note that the total 
contribution of satellites is not sufficient to alter the apparent star formation mode (i.e., 
main-sequence or starburst) of the centrals as determined from the \emph{Herschel}/PACS data. 
Finally, we can add to this value the SFR derived from the uncorrected rest-frame UV. Using the 
$B$-band flux as a measure of the rest-frame 1500$\AA$ emission, this yields $14\pm3$ and 
$8\pm1$~M$_{\odot}$~yr$^{-1}$ for the satellites and centrals, respectively. The total SFR of 
satellites estimated from FIR and uncorrected UV is then $60\pm7$~M$_{\odot}$~yr$^{-1}$, or 
$\sim33$\% of the central's.\\

\subsection{\label{xray}X-ray observations}

\begin{figure}
\centering
\includegraphics[width=0.4\textwidth]{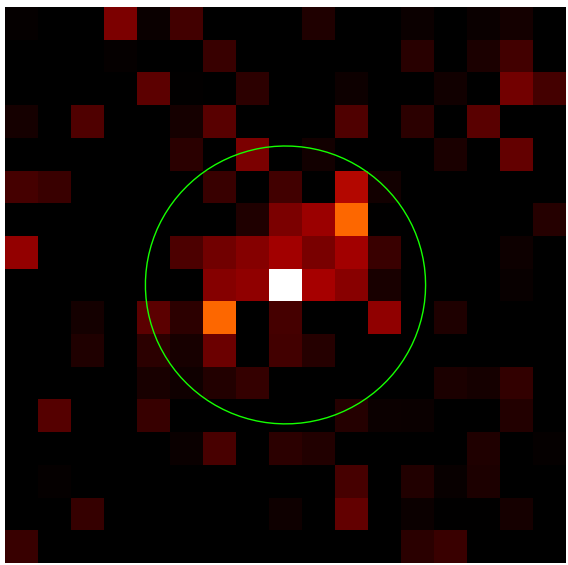}
\caption{Stacked X-ray image at the position of the centrals (excluding individual detections), 
from the combined \emph{Chandra} and \emph{XMM-Newton} $0.5-2$~keV image of the COSMOS field. 
For reference, the green circle has a radius of 15\arcsec.}
\label{fig:x}
\end{figure}

Since our sample is slightly different and smaller than the one used in \citet{Bet14}, the average 
mass of the host halos studied here might be different from that reported in that paper. As previously, 
we have used deep X-ray observations of the COSMOS field by the \emph{Chandra} and \emph{XMM-Newton} 
observatories \citep[see, e.g.,][]{Fin07,Elv09} to constrain the total mass of the halos through a 
stacking analysis. 
Of the 215 centrals in the sample, 14 are directly detected as extended sources and 152 are located in 
zones free from emission. Most of the direct detections appear to be consistent with chance associations 
along the line of sight with lower-redshift galaxy groups, including 3 that were already known 
\citep{Geo11}. We have accordingly excluded the ``direct detections'' from the X-ray stack.  However, 
keeping or removing these objects from the sample has no effect on the rest of the analysis presented 
in this paper and its conclusions. For the sources in regions free from detectable emission, we have 
used the background-subtracted and exposure-corrected X-ray image, after subtracting detected point 
sources. The average flux in the 0.5--2~keV band is then $1.1\times10^{-16}$~erg~cm$^{-2}$~s$^{-1}$, 
detected at 5.3$\sigma$. Fig.~\ref{fig:x} shows a stacked image of these individually undetected objects. 
For halos in the range $z\sim1.5-2.5$, using the calibrations of \citet{Leau10}, this flux corresponds 
to a rest frame 0.1--2.4 keV luminosity of 0.8--2.9$\times 10^{43}$~erg~s$^{-1}$, an intergalactic medium 
temperature of $\sim 1$~keV and a total mass of M$_{200}=2.1-2.4\times10^{13}$~M$_{\odot}$, values similar 
to those reported in \citet{Bet14}. Such sources might then soon be individually detectable
in deeper X-ray surveys such as the CDF-S \citep{Fin14}, where the applicability of the \citet{Leau10} 
scaling relations has already been verified for sources with fluxes close to that reported here.

\subsection{\label{hod}Comparison with model predictions}

\begin{figure}
\centering
\includegraphics[width=0.45\textwidth]{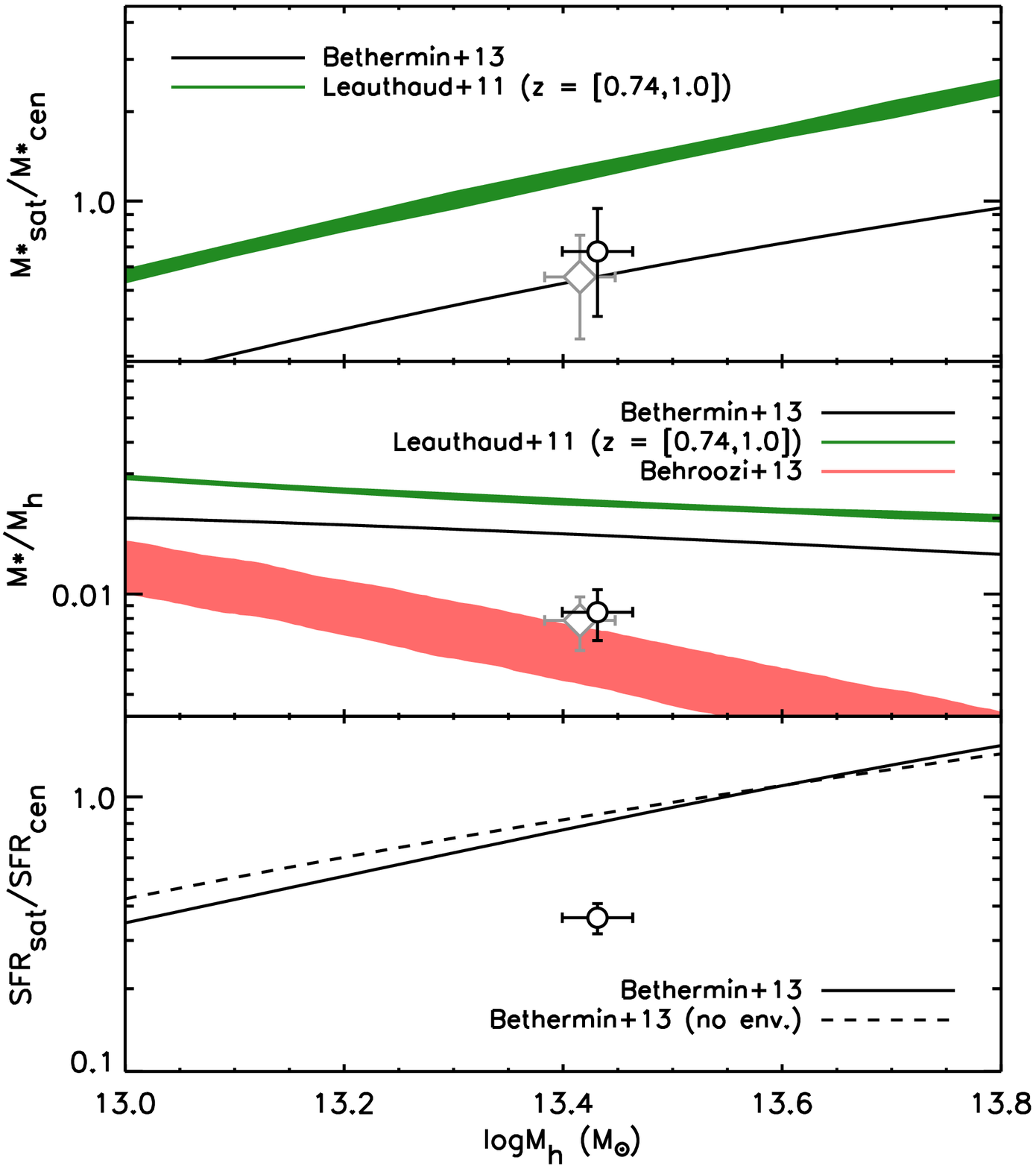}
\caption{Comparison of the integrated properties of halos in our sample, as a 
function of halo mass, with model predictions from \citet[black]{Bet13}, \citet[green]{Leau12}, 
and \citet[red]{Beh13}: stellar mass fraction of satellites with respect to the central 
(top), stellar to total mass ratio (middle), and SFR fraction of satellites compared to that of 
the central (bottom; the dashed line corresponds to a case without environmental effects). 
The black circle and gray diamond show, respectively, the MF-corrected and uncorrected values. 
The latter have been shifted to the left for clarity. 
}
\label{fig:hod}
\end{figure}

We have compared the integrated mass and SFR of satellites to the predictions of different 
halo occupation distribution (HOD) models from \citet{Leau12}, \citet{Beh13}, and 
\citet[hereafter, respectively, L12, Bh13 and Bt13]{Bet13}. 
The L12 model is shown for its highest defined redshift bin ($z=0.74-1$) while the last two 
are both evaluated at $z=1.8$.
Fig.~\ref{fig:hod} shows this comparison for three quantities: the ratio of the total stellar mass 
of satellites to that of the central, $M_{sat}/M_{cen}$, the fraction of stellar mass (central 
and satellites) to total mass, $M_{\star}/M_h$, and the ratio of the total SFR of satellites to the 
SFR of the central, SFR$_{sat}/$SFR$_{cen}$. We have here used the total halo masses derived from X-ray 
stacking, MF-corrected stellar masses and UV+FIR SFRs. All quantities assume the same IMF.

The contribution of satellites to the total stellar mass is consistent with the predictions 
of Bt13 and below L12 at $z\lesssim1$ (this is to be expected, since $M_{sat}/M_{cen}$ 
decreases with increasing redshift in both models). We note that this would still be the case if 
only the measured, uncorrected satellite mass were used. 
The mass fraction of satellites also appears to be compatible with that predicted at $z=0$ 
by recent numerical simulations, in the case of $\sim10^{13}$~M$_{\odot}$ halos \citep{Gen14,Kra14}. 
In our probed mass range, the stellar mass of the central does not evolve much with redshift 
\citep{Mos13,Beh13}, but it is not obvious that this should also be the case for the stellar 
content of satellites. This might therefore suggest that the processes that determine 
the baryon conversion efficiency of the host halo also determine to the first order that of 
the sub-halos. These processes can either act within the host halo (what is usually thought of 
when considering environmental effects), or in the large-scale structure containing both the 
central and satellites. In the second case, they would then be related to the conformity of galaxy 
properties on large scales seen at low redshift \citep[e.g.,][]{Par08,Ann08,Kau10,Kau13} and help 
synchronize the stellar mass build-up of the central and its future satellites, before the 
satellites merge with the host halo.\\

On the other hand, the measured total stellar mass fraction $M_{\star}/M_h$ is somewhat lower 
than the predictions of Bt13 and more consistent with Bh13. This is not entirely surprising, 
as the former is optimized to reproduce FIR counts while the latter adopts a more sophisticated 
treatment of the stellar-to-halo mass relation. The two models also use different stellar mass 
functions. Furthermore, Fig.~\ref{fig:hod} does not include the systematic uncertainty 
on stellar mass estimates ($\sim0.2-0.3$~dex; see also Section~\ref{data}). If we take it into 
account, both the Bt13 and L12 models become compatible with the measured value.
Finally, the derived SFR ratio of satellites and centrals is substantially lower than model 
predictions. This might seem surprising, since the mass ratio is itself fully consistent 
with expectations. 
On the other hand, the total SFR of satellites is, in this model, somewhat sensitive to both 
the behavior of the MF at low masses and the slope of the main sequence. For example, if we 
assume a slope of unity, instead of the value of 0.8 used by Bt13, the predicted SFR ratio 
would decrease by a factor of $\sim2$, making it more consistent with observations. The Bt13 model 
also adopts a relatively simplified treatment of star formation in sub-halos: in the 
``no-environment'' case, the SFR and the quenched fraction are both a function of sub-halo mass, 
while in the other case the model assumes that all satellites of active centrals are themselves 
active. Notably, suppressed (but non-zero) star formation and gradual quenching are not considered.

\section{\label{sfr}Satellite properties as a function of radius}

\begin{figure*}
\centering
\includegraphics[width=0.49\textwidth]{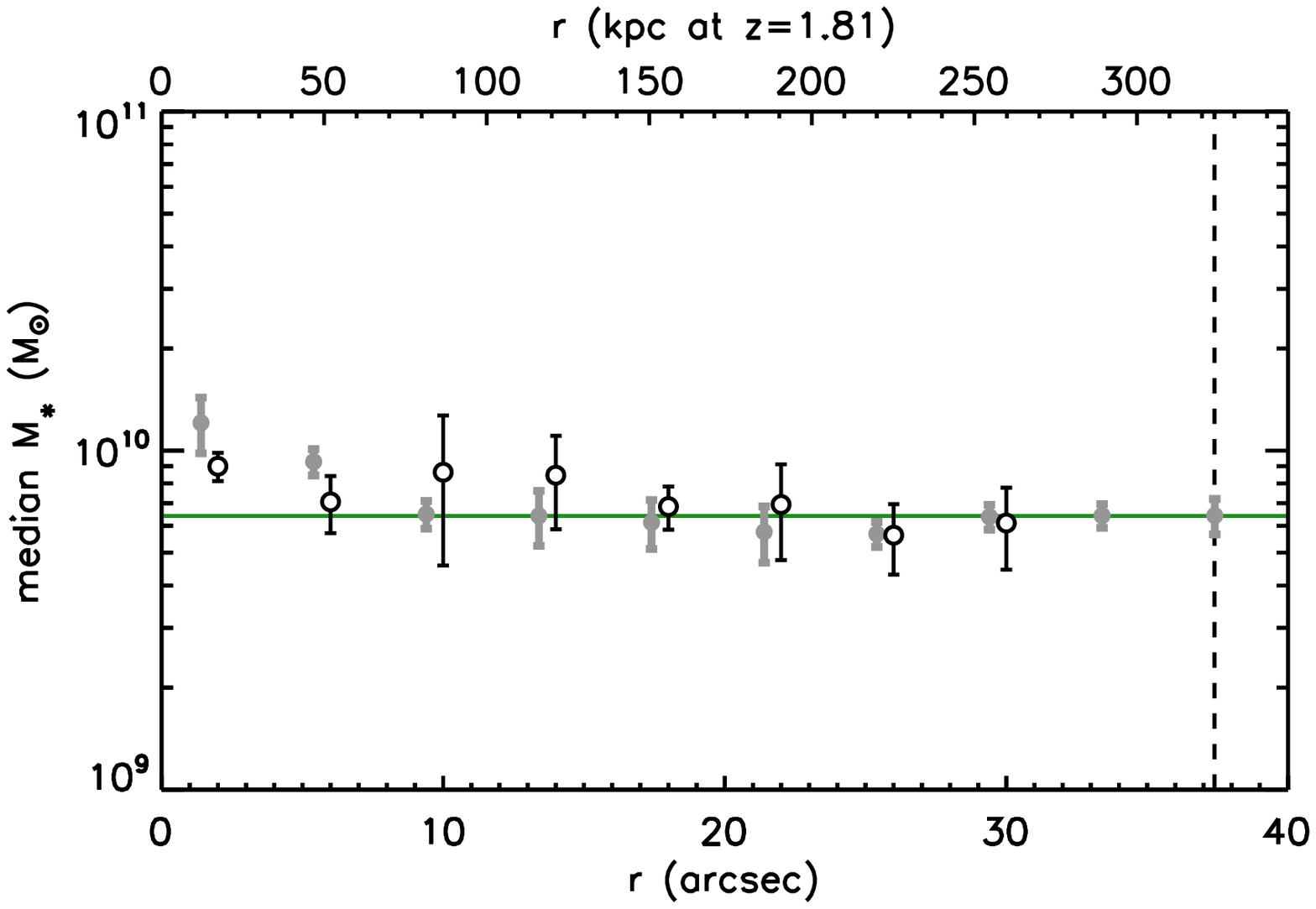}
\includegraphics[width=0.49\textwidth]{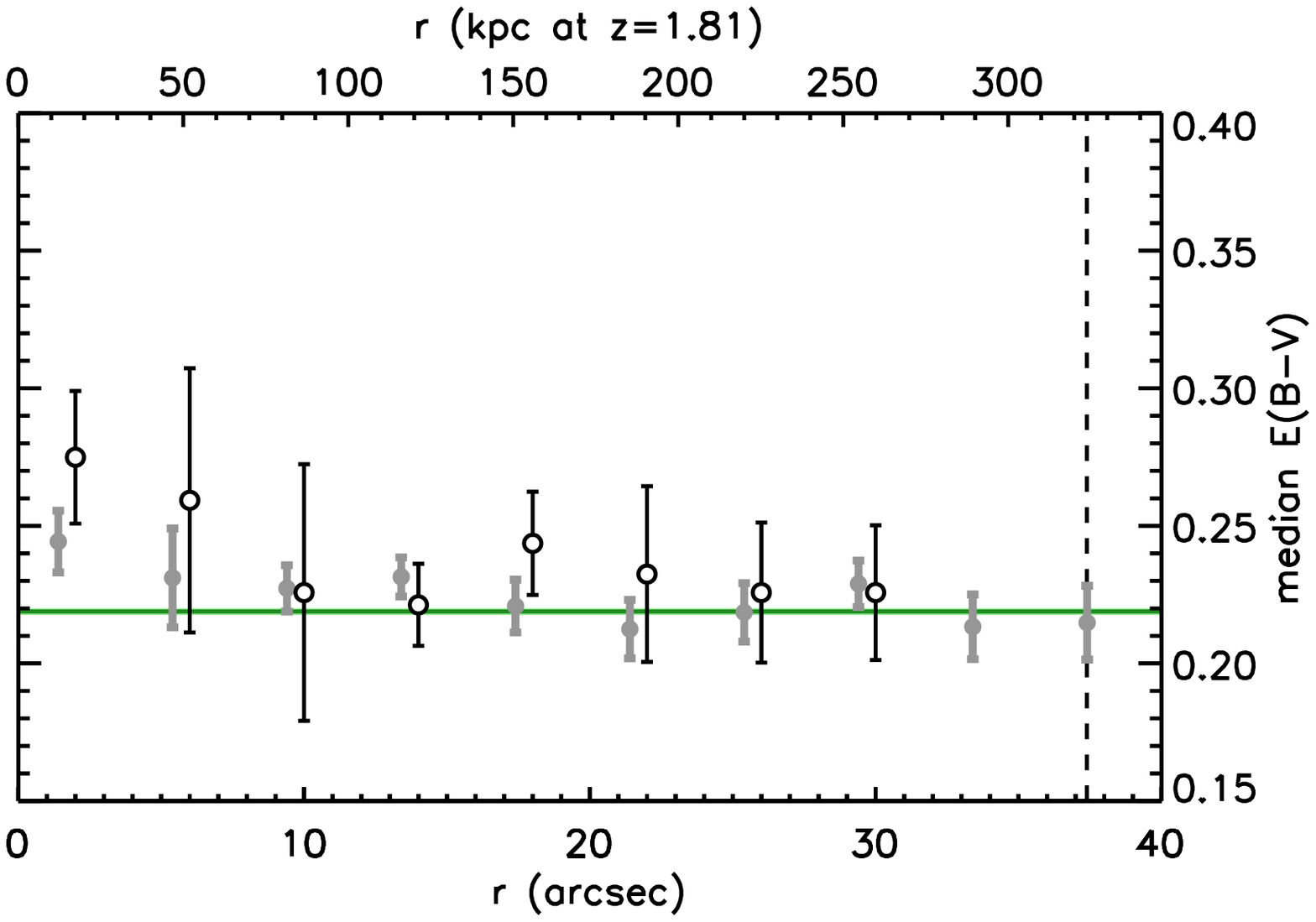}
\includegraphics[width=0.49\textwidth]{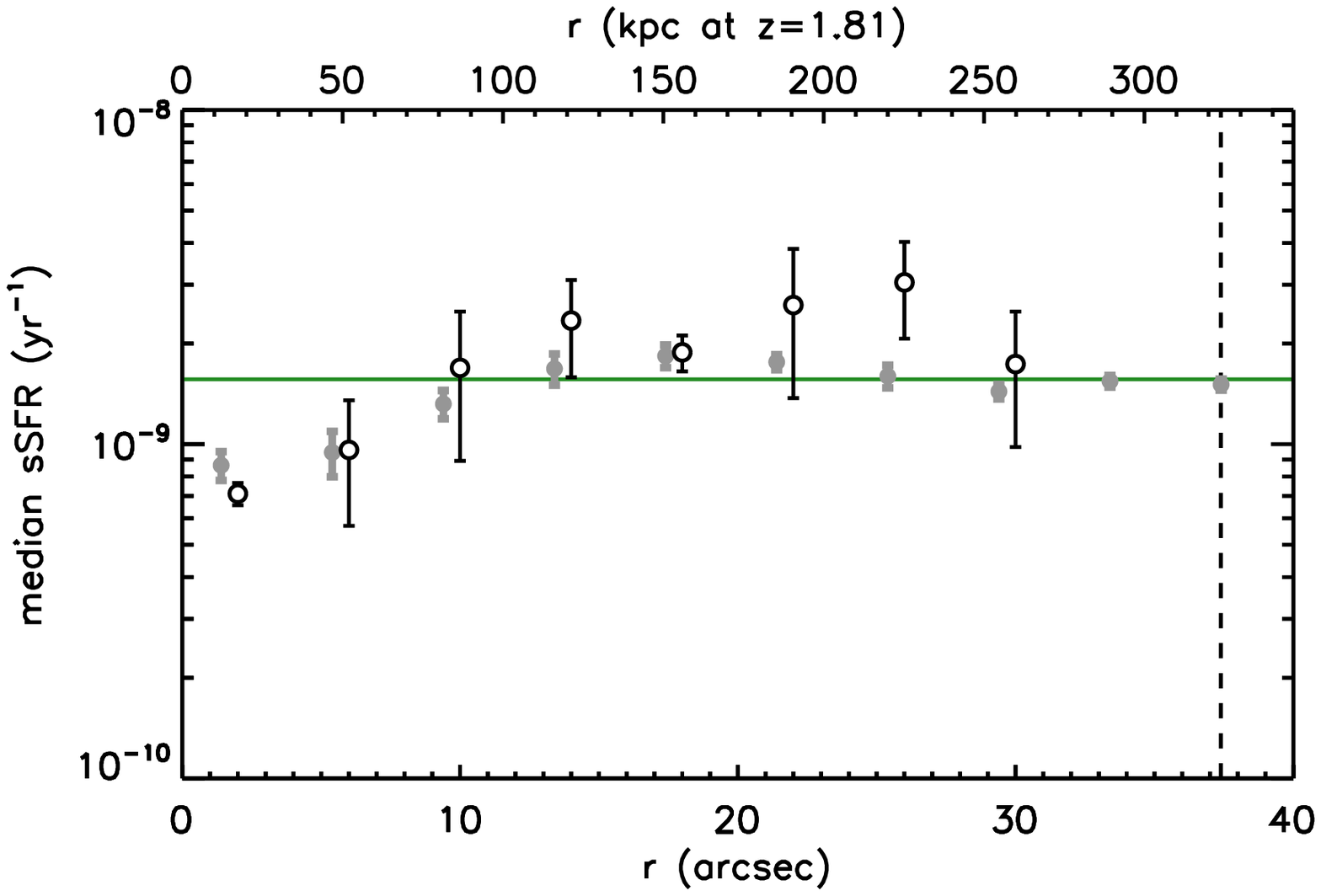}
\includegraphics[width=0.49\textwidth]{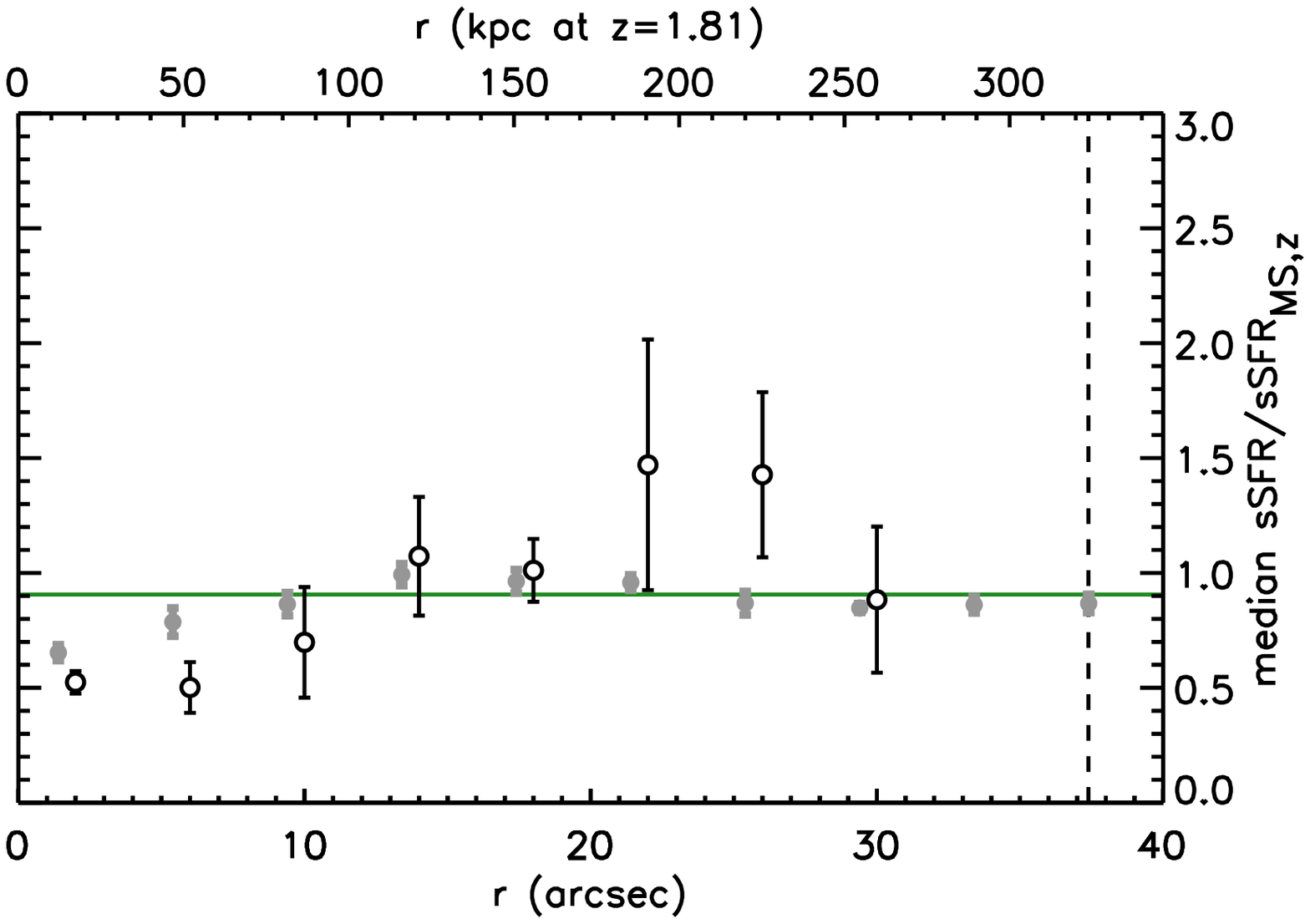}
\caption{Median background-subtracted stellar population parameters of UVJ-selected 
star-forming satellites (open symbols), as a function of distance to the central, in bins of 
4\arcsec: stellar mass (\emph{top left}), extinction (\emph{top right}), sSFR (\emph{bottom 
left}), and sSFR as a fraction of the main-sequence value (assuming a slope of 0.8). 
Median values for the total distribution of star-forming satellites (i.e., without subtracting 
the background distribution) are shown in gray, slightly shifted to the left for convenience. 
Background levels and related uncertainties, estimated at $r>50$\arcsec, are shown in green 
and the virial radius is indicated by a dashed line.
}
\label{fig:ssfr}
\end{figure*}

In this section, we investigate the variation of the stellar population properties of 
star-forming satellites with distance to the central. As in Section~\ref{stack}, 
we have selected star-forming satellites based on their rest-frame UVJ colors, using 
the high-redshift criterion of \citet{Wil09}. Conservatively, we have also excluded 
nominally star-forming objects that are within 0.1~mag of the dividing line, so as 
to avoid possible contamination from quiescent satellites. Fig.~\ref{fig:ssfr} 
shows the radial dependency of dust extinction, stellar mass, and specific star 
formation rate (SFR/M$_{\star}$, or sSFR). We here look at the variation of median 
values to minimize the effects of outliers. However, because we can expect $\sim20$\% of 
spurious associations even in the central b in (see Fig.~\ref{fig:pos}), this measure could 
still be skewed by interloper contamination. To mitigate this, we performed, for each measured 
quantity, the following statistical background subtraction: in each radial bin within $r_{vir}$, 
we randomly removed a number of satellite candidates corresponding to the expected number of 
interlopers, using the background distribution as prior. The uncertainties were estimated from the 
dispersion of median values of these background-subtracted distributions. To these values, we have 
added, as in Section~\ref{stack}, the uncertainties derived from bootstrap resampling. 
This subtraction was performed up to the putative virial radius, although the satellite 
counts start becoming consistent with background levels already at $r\gtrsim20$\arcsec 
(or $\sim170$~kpc; see Fig.~\ref{fig:pos}). 
For comparison, Fig.~\ref{fig:ssfr} also shows the median value prior to the statistical 
subtraction.\\

The stellar mass and extinction of satellites does not vary very much with radius, except 
in the central bins, where the median M$_{\star}$ and $E(B$-$V)$ are higher than the 
background value by $\sim0.2$~dex and $\sim0.04$~mag, respectively. 
More surprisingly, while the SFR density of satellites increases monotonously with decreasing 
halo-centric distance (Fig.~\ref{fig:prof}), their median sSFR varies significantly within the 
halo, first exhibiting a mild ($\gtrsim20$\%, $\sim3\sigma$ with respect to the background value) 
rise at 0.5$r_{vir}$ ($\sim150$~kpc), then a more significant decrease ($\sim40$\%, $5\sigma$) 
close to the central ($<50$~kpc). Several interpretations, which we discuss below, could account 
for this effect.\\

\emph{Normal sSFR variation}, as a consequence of the higher median mass of satellites at the 
center since, in the case of a non-unity slope for the SFR-$M_{\star}$ relation, the sSFR is 
mass-dependent. 
However, assuming a slope of $\sim0.8$ \citep{Rod14} and considering the ratio of the sSFR of 
individual galaxies to that of the main sequence at their stellar mass does not decrease the 
significance of the sSFR drop at small radii and only slightly that of the excess at 
$100-200$~kpc (from 3$\sigma$ to 2$\sigma$), as shown in Fig.~\ref{fig:ssfr} (bottom right).
In fact, a very shallow slope of $\sim0.3$ would be needed to fully account for the observed 
sSFR decrease. This value seems unlikely for $\sim10^{10}$~M$_{\odot}$ galaxies at $z\sim2$ 
\citep[even in the case of a broken power-law relation;][]{Whi14} and, in the case of a 
single power-law, would reinforce the excess at $\gtrsim150$~kpc. On the other hand, a slope 
value of near-unity would not alter the shape of the sSFR variation.\\

\emph{Stellar population modeling bias} or \emph{``missed'' SFR} from heavily obscured star 
formation, e.g., due to a systematic underestimation of the extinction-corrected SFR in redder 
galaxies. 
We have performed a set of simulations using our stellar population models with varying extinctions 
and S/N ($E(B$-$V)=0-1$ and S/N$\geq3$, respectively) to test the first possibility and quantify 
the bias to stellar population properties in our SED fitting. 
We find that, when increasing the extinction, faint objects will tend to have their stellar mass 
underestimated by $\sim0.05$~dex and their reddening and SFR overestimated by $\sim0.02$~mag 
and $\lesssim10$\%, respectively. These values are within the uncertainties of their respective 
parameters. This is not very surprising, as extinction-corrected SFRs derived from UV-NIR SEDs 
have already been found to be quite robust \citep[e.g.,][]{Rod14} and the COSMOS field benefits 
from a large multiwavelength coverage.\\ 

\begin{figure*}
\centering
\includegraphics[width=0.49\textwidth]{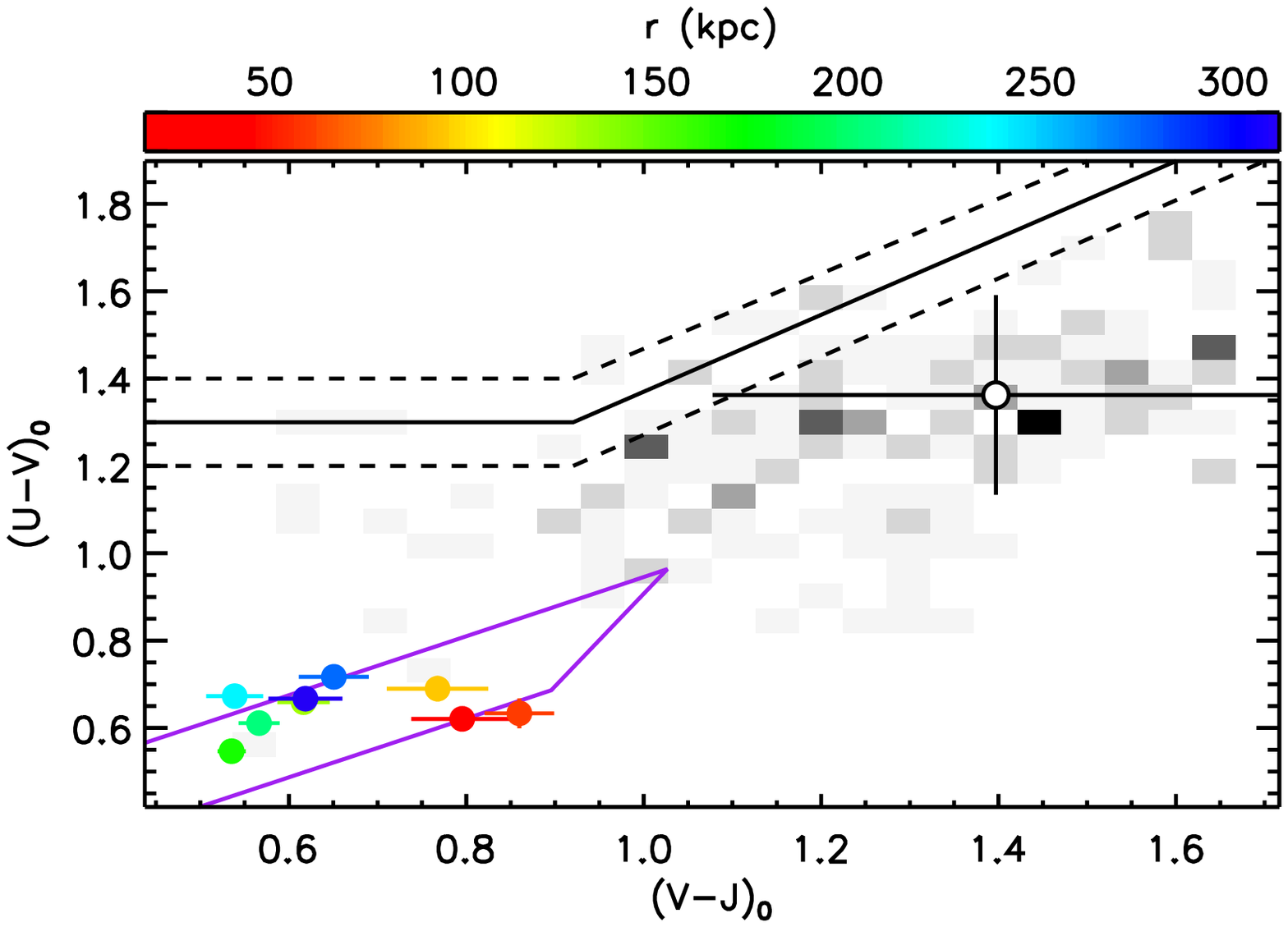}
\includegraphics[width=0.49\textwidth]{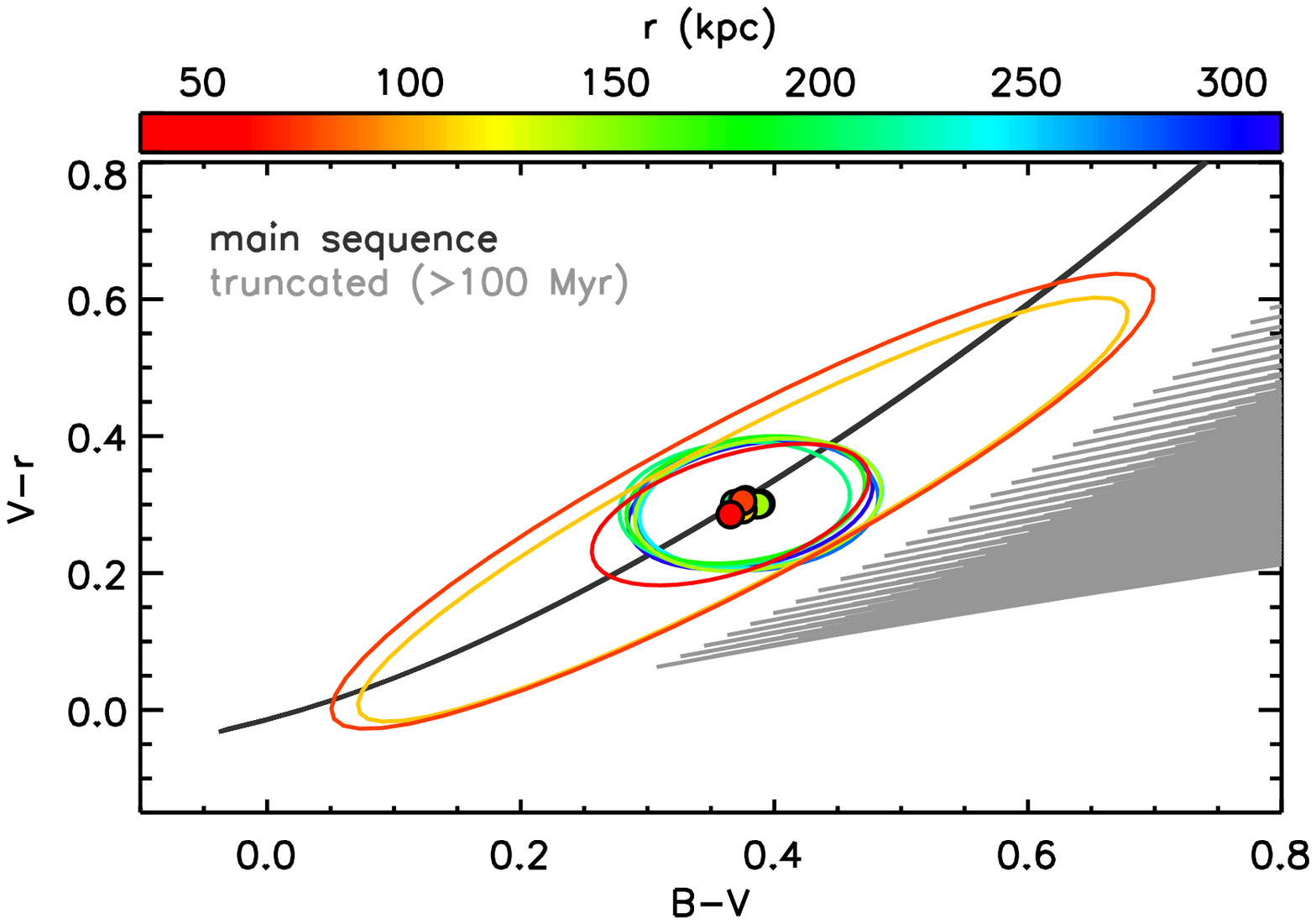}
\caption{\emph{Left}:~Median rest-frame $U$-$V$ and $V$-$J$ colors, derived from SED 
modeling, of star-forming satellites (filled circles, colored as a function of distance 
to the central) and centrals (open circle and gray density histogram). The purple lines 
show the expected colors of stellar populations models of varying age following a 
``main-sequence'' SFH, assuming $E(B$-$V)<0.3$. 
\emph{Right}:~median observed $B$-$V$ and $V$-$r$ colors of satellites, as a function of 
distance to the central, compared to the expected colors of a star-forming model 
(main-sequence SFH, black) and a constant SFR one observed $\geq100$~Myr after the 
quenching of star formation (gray). We here include dust extinction with $E(B$-$V)=0-2$. 
The ellipses show the dispersion of $B$-$V$ and $V$-$r$ values in each radial bin. 
}
\label{fig:uv}
\end{figure*}

In extreme cases heavily obscured star-forming regions in galaxies could be missed 
entirely by UV-based estimates. In this scenario, Fig.~\ref{fig:ssfr} could then 
be interpreted as implying a change of star formation mode in satellites as they fall 
closer to the central, toward heavy obscuration. This would result in a systematic 
underestimate of the SFR at small radii. Such obscured star formation would however still 
contribute to the integrated rest-frame near- and far-IR light of the galaxies, and its 
influence be detectable in broad-band photometry. As shown in Fig.~\ref{fig:uv} (left), 
satellites closer to the central galaxy do indeed appear to have slightly redder ($\sim0.2$~mag) 
rest-frame $V$-$J$ colors than field galaxies, although the two populations still have 
compatible $U$-$V$ values and remain within the locus of low-extinction star-forming 
galaxies. This color difference could be due to a combination of factors, such as longer 
star formation timescales or higher metallicities (both would increase $(V$-$J)_0$ but 
decrease $(U$-$V)_0$) in association with higher ages or extinction (which increase both 
$(V$-$J)_0$ and $(U$-$V)_0$).
On the other hand, the excellent agreement of the FIR SED of satellites with 
main-sequence models, and between the FIR and SED-derived total SFRs, suggests that 
``hidden'' star formation is not present in significant quantities 
\citep[see also][for similar conclusions]{Zan15}.\\

\emph{Environmental effect} on the activity of satellites, from interaction with the 
halo and/or the central. In this case, the observed sSFR decrease could originate from 
two different galaxy populations: systems with non-zero but suppressed star formation, 
and galaxies where it has recently ceased altogether. 
As the rest-frame UV bands we have used to select and characterize star formation 
directly trace the light of massive stars, our estimates are only sensitive to 
timescales in excess of $100$~Myr (e.g., a galaxy can be expected to stay 
in the star-forming locus of the UVJ plane for $\sim300-500$~Myr after cessation 
of star formation). However, recently quenched systems can be easily distinguished 
from their still active counterparts due to the aging of their stellar population 
affecting bluer bands first. Fig.~\ref{fig:uv} (right), for example, shows two rest-frame 
UV colors of satellites compared to young stellar population models with and without ongoing 
star formation. At $z\sim1.8$ the $B$, $V$, and $r$ filters sample the $1500-2300\AA$ 
rest-frame and are thus very sensitive to UV light from short-lived massive stars.
We see no correlation between the shape of the UV continuum and distance from the central, 
with satellites at all radii being on average consistent with ongoing star formation.\\

\begin{figure}
\centering
\includegraphics[width=0.49\textwidth]{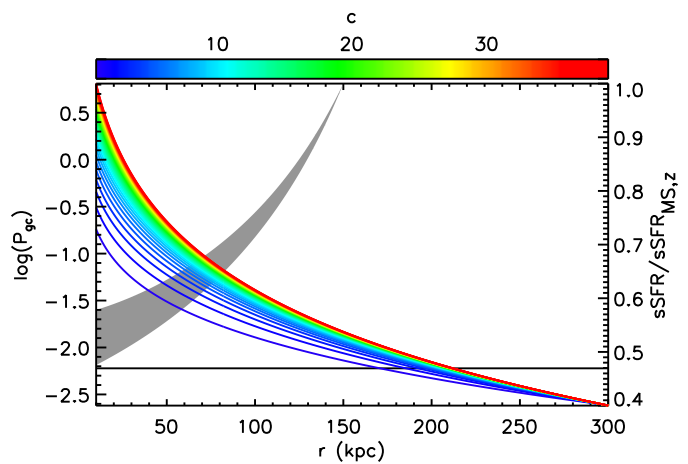}
\caption{Decrease in sSFR in a simple no-infall, no-recycling case as a function of distance 
to the center of the halo (gray, right axis; arbitrarily starting at $r=150$~kpc) and tidal 
perturbation parameter $P_{gc}$ (red to blue, left axis), for a low-mass satellite. The 
horizontal line shows the threshold value at which the gravitational influence of the halo 
can trigger inflow within satellite galaxies \citep{BV90}. Both quantities are shown for NFW 
concentration parameters ranging from $c=1$ to $c=40$.
}
\label{fig:ff}
\end{figure}

We therefore conclude that the observed sSFR decrease in UVJ-selected star-forming 
satellites reflects an actual depression of star formation induced by the group 
environment. We can estimate a lower limit on the timescale of this effect 
in the following way: in a pure free-fall case, a galaxy along a radial orbit 
starting at $r_{vir}$ would reach the center of the halo after 
$\sim600-700$~Myr (depending on the concentration of the total matter distribution, 
if we assume a NFW profile), reaching velocities of $\sim1000-2000$~km~s$^{-1}$ and 
needing only $\sim100-150$~Myr to cross the last 150~kpc, i.e., the radius corresponding 
to the observed decrease of sSFR. If we assume, to the first order, that the sSFR drop is due 
to the absence of gas accretion from the satellites' reservoirs, and that recycling plays a 
negligible role, we find \citep[following][]{Erb08} that the time required for the sSFR 
to decrease to the observed level would indeed be $\lesssim150$~Myr. 
This is illustrated in Fig.~\ref{fig:ff}, where we plot, as a function of radius, the 
diminution of sSFR, assuming the satellites experience no gas infall at $r<150$~kpc. 
A more circular orbit would increase the time spent by the satellites interacting with 
the inner halo, while a more gradually diminishing gas supply (as well as some recycling) 
would also increase the quenching timescale. 

Several mechanisms can affect the gas reservoirs of galaxies in dense environments 
and induce a diminution of star formation \citep[for a review, see, e.g.,][]{Bos06,Par09}. 
Interactions between satellites are here likely not a significant driver of galaxy evolution, 
as the galaxy density around individual centrals is relatively low. The minimum separation 
of satellites is $\sim70-90$~kpc in projection closest to the central (where the signal is 
dominated by real satellites rather than interlopers; see Fig.~\ref{fig:pos}), an order of 
magnitude larger than the typical galaxy size in this redshift and mass range 
\citep[e.g.,][]{vdW14}, and already above the scale at which galaxy ``harassment'' is 
effective \citep{Moo96}. Because of the redshift uncertainties for individual satellites, 
we can expect that the actual distance between them be significantly higher. 
On the other hand, interaction with the hot diffuse intra-halo gas, whose presence 
is confirmed by X-ray stacking, constitutes a more plausible source of 
environmental forcing. The hot gas medium can efficiently shut down star formation, 
mostly through hydrodynamical interaction, by either preventing further accretion of 
cold gas onto the galaxies \citep[e.g., ``starvation'',][]{Lar80,Bek02} or through 
outright stripping of the galaxies' interstellar gas \citep{Gun72,Nul82}. 
These mechanisms are commonly invoked to explain general properties of galaxy 
populations in clusters, such as systematic sSFR differences with respect to 
field galaxies \citep[e.g.,][]{vdL10,Al14} \emph{and} the lack thereof. In particular, 
in massive, high-redshift clusters the sSFR of star-forming galaxies does not appear to be 
much correlated with cluster-centric distance \citep[e.g.,][]{Muz12}. This, together with 
the phase-space distribution of different galaxy populations \citep{Muz14}, is viewed as 
a sign that the quenching of star formation in dense environments happens on short 
timescales. 
The systems studied here probe not only a somewhat higher redshift range than the 
aforementioned studies, comparable in fact to the current limit of massive 
cluster samples, but also a mass range that is an order of magnitude lower. They are 
dynamically simpler than large clusters and with lower velocities, gas temperatures, 
and densities. The interactions of satellites with their environment should then be less 
violent. Longer interaction timescales might thus explain the apparent discrepancy between 
our analysis, which finds a clear sSFR trend, and cluster studies, where such an effect 
is not seen. 
On the other hand, in the limit case described above (radial orbit, no gas infall at 
$\lesssim150$~kpc), a galaxy falling toward the halo center would have its sSFR decrease 
by 1~dex in $\sim1$~Gyr, corresponding to an $e$-folding time of $\sim0.3$~Gyr. This short 
timescale is similar to that inferred for massive clusters and consistent with a fast 
quenching \citep[see also, e.g.,][]{Wetz13}. 
While constraining the actual mechanisms acting on the satellites is beyond the scope of 
this paper and of the data, we note that such timescale is still consistent with either 
classical ``starvation'' \citep[i.e., mechanical stripping of the gas reservoir;][]{Bek02} 
or shock heating of the gas, as predicted by hydrodynamical simulations 
\citep[here, the interaction between satellites and their host halo would happen at $z<2.5$ 
in all cases, at an epoch when halos of $\gtrsim10^{13}$~M$_{\odot}$ are expected to be hot 
and thus prevent efficient cooling of the gas;][]{Dek06}. 
On the other hand, \citet{Zip13} report no such sSFR gradient in lower redshift groups of 
similar mass. This might reflect a difference between $10^{13}$~M$_{\odot}$  halos at $z\sim2$ 
and $z\sim1$, in timescales for environmental processes or of baryon content. We note however 
that their highest redshift bin ($1.2<z<1.7$) shows a hint of a $\sim0.4$~dex sSFR drop similar 
to what we report here, although it is not significant enough due to the bin containing only 
one object of uncertain nature \citep{Kur09}.\\

On the other hand, the processes described above cannot account for the observed sSFR excess 
in satellites at $\sim150$~kpc. We can discount galaxy-galaxy interactions for the same reasons, 
the median minimum separation of satellites in individual halos being even higher at $>100$~kpc 
and indistinguishable from field levels. Tidal interaction with the halo, however, by 
perturbing the gas already in the galaxies, could accelerate star formation in satellites and 
contribute to clearing their gas \citep[see][at similar redshift]{Val15}. 
Following \citet{BV90}, we estimate the tidal perturbation parameter 
$P_{gc}=(M_h(r)/M)/(r/A)^3$, where $M$, $A$, $r$ and $M_h(r)$ are, respectively, the mass and 
radius of the satellite, its distance from the halo center and the halo mass enclosed within 
$r$. For $M_h(r)$ we assume a NFW profile with a varying concentration in the range $c=1$--40 
and a total mass given by the X-ray estimate. We find that in our case the tidal perturbation 
starts becoming significant ($P_{gc}>0.006$, assuming no stabilizing stellar halo) at a distance 
of $r<200$~kpc from the halo center, assuming a characteristic satellite mass of 
$M=7\times10^{9}$~M$_{\odot}$ (see Fig.~\ref{fig:mz}) and size of $A\sim3$~kpc \citep{vdW14}.\\

\subsection{\label{qf}Passive fraction}

\begin{figure}
\centering
\includegraphics[width=0.49\textwidth]{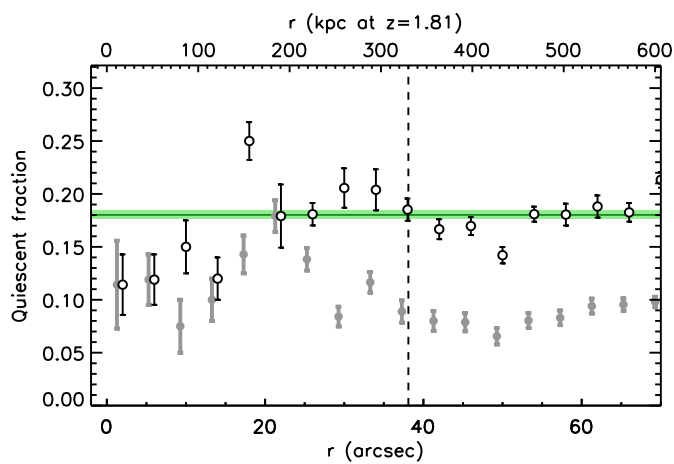}
\caption{Fraction of quiescent satellites as a function of radius. The satellites were selected 
according to the UVJ criterion (open symbols), and adding a margin of 0.1~mag in both $(U$-$V)_0$ 
and $(V$-$J)_0$ (gray symbols). Values within the virial radius, as shown by the dashed vertical 
line, were derived from background-subtracted distributions using the same procedure as in 
Fig.~\ref{fig:ssfr}. 
Background levels and related uncertainties, estimated at $r>50$\arcsec, are shown in green. 
}
\label{fig:qf}
\end{figure}

Finally, we note that the observed sSFR decrease at small radii is not mirrored by an increase 
in the number of quiescent satellites near the central. In Fig.~\ref{fig:qf}, we show the ratio 
of UVJ-selected quiescent satellites to the total number of satellites in each radial bin. 
We performed the same statistical background subtraction as described above, using the color 
distributions of the satellites as priors and estimating the quiescent fraction for each 
random trial. The fraction of quiescent galaxies is close to 20\% at large radii and appears 
to decrease slightly at $r<20$\arcsec. This value and trend are similar to those derived 
by \citet{Har15} from a slightly lower redshift sample. On the other hand, if we adopt a 
slightly more stringent criterion, by adding a 0.1~mag margin (see Fig.~\ref{fig:uv}) and 
selecting only the redder UVJ-quiescent galaxies, the background quiescent fraction drops to 
$\sim$10\% and the trend at small radii disappears. This suggests that, at least in our case, 
it is mostly due to objects close to, or straddling, the dividing line between the passive and 
star-forming loci. The stellar mass distributions of both quiescent samples are not significantly 
different, however. 
The absence of a clear number excess could seem counterintuitive, considering the sSFR 
variation described in Section~\ref{sfr}. On the other hand, the appearance of an obvious 
quiescent galaxy population takes time. For it to happen in this case, the quenching of star 
formation in the central satellites would have to have started at $z\sim2.5$, if we assume a time 
span of 1~Gyr for a 1~dex sSFR decrease as discussed above. This would in turn imply that the 
environmental conditions responsible for it (e.g., a hot halo) be already in place at this 
epoch. We can infer that this was not the case in the type of halos investigated here.

\section{\label{conc}Conclusions}

Low-mass structures traced by massive galaxies, while more difficult to confirm individually, 
can be efficiently selected statistically. At high redshift, they can offer a more accessible 
window to galaxy evolution in dense environments than galaxy clusters, their high abundance 
compensating for the lower galaxy number density and environmental bias, even in a relatively 
limited area. We have here taken advantage of the wealth and depth of photometric data available 
on the COSMOS field to study the distribution and properties of star-forming satellites associated 
with massive galaxies on the main sequence of star formation, as tracers of group-size halos of 
mass $\sim2-3\times10^{13}$~M$_{\odot}$. We have constructed a sample of massive star-forming 
galaxies at $\langle z\rangle =1.8$, selecting only objects without close neighbors of comparable 
mass so that they be putatively central to their host halo. 
We have verified the average total mass of said halos thanks to deep \emph{Chandra} and XMM data, 
and found it to be $\lesssim3\times10^{13}$~M$_{\odot}$. 
Using the recently released matched photometric catalogs for the COSMOS field, we have derived 
stellar population parameters for both centrals and satellites. Our conclusions are the following: 

\begin{itemize}

\item we have estimated the contribution of satellite galaxies to the stellar mass 
and SFR of the systems at, respectively, $\sim68$\% and $\sim25-35$\% of the stellar mass and 
SFR of the central galaxy (or $\sim40$\% and $\lesssim25$\% of the total stellar mass and SFR), 
after correcting for the completeness limit of the sample. The stellar mass fraction of satellites 
with respect to the central is found to be consistent with the predictions of HOD models, as 
is the total stellar mass to halo mass ratio. On the other hand, the observed total SFR of 
satellites appears to be a factor of $\sim2-3$ lower than model predictions. This might be related 
to the relatively simple treatment of star formation in sub-halos adopted by our chosen model, 
or to assumptions on the behavior of the main sequence of star formation at low stellar mass.

\item we have also independently estimated the SFR of satellites and centrals from stacked FIR 
data, by separating their contributions through source decomposition. The SED thus derived is 
well-fitted by a main-sequence template and yields a SFR of $\sim47$~M$_{\odot}$~yr$^{-1}$, 
consistent with the UV-NIR estimate. This also suggests an absence of significant heavily 
obscured star formation (e.g., starbursts) in the satellite population.

\item finally, we have probed the radial dependence of the properties of star-forming satellites. 
We find significant variation of their sSFR within the virial radius, with a marginal excess at 
$r\sim150$~kpc followed by sharper drop at $r<100$~kpc. This suggests that the group environment 
acts differently on star-forming galaxies within $r_{vir}$ depending on their distance to the 
center, enhancing star formation slightly at larger radii while quenching it with a timescale of 
$\gtrsim300$~Myr closer to the center. In the first order, this is consistent with destabilization 
of galactic gas by the halo potential followed by prevention of further gas accretion, as the 
galaxy falls closer to the center of the halo.
\end{itemize}

On the other hand, the use of photometric data not only implies some amount of back- and foreground 
contamination, but also precludes knowledge of important quantities, such as the instantaneous star 
formation rate and metallicity, that would more precisely constrain the mechanisms of galaxy evolution 
in these halos. 
Wide-field, high-coverage spectroscopic instruments (e.g., large integral field units such as MUSE) 
and, later, ``all-in-one'' large-scale surveys (e.g., \emph{Euclid} and WFIRST), should allow for a 
dramatic improvement in statistics and redshift resolution, especially around the critical epoch of 
galaxy and cluster progenitor build-up at $z\sim2$.

\begin{acknowledgements}

RG, ED, MB, MS, VS and FV were supported by grants ERC-StG UPGAL 240039 and ANR-08-JCJC-0008.\\ 

\end{acknowledgements}

\end{document}